\documentclass[pra,aps,twocolumn,showpacs,superscriptaddress,nofootinbib,floatfix]{revtex4-2}
\pdfoutput=1
\usepackage{bbold}
\usepackage{amsmath}
\usepackage{mathtools}
\usepackage{graphicx}

\usepackage{multirow}
\usepackage[table,xcdraw]{xcolor}
\usepackage[colorlinks=true, allcolors=blue]{hyperref}

\newcommand{\Graph}{{\mathcal{G}}}
\newcommand{\edges}{{\mathcal{E}}}
\newcommand{\vertices}{{\mathcal{V}}}

\newcommand{\ket}[1]{\left|#1\right\rangle}

\newcommand{\argmax}{\text{argmax}}
\newcommand{\argmin}{\text{argmin}}

\usepackage{amsmath,amssymb}
\usepackage[a4paper, portrait,text={19cm,26cm},centering]{geometry}
\usepackage[most]{tcolorbox}
\usepackage{mathrsfs}
 \usepackage{graphicx}
\newtheorem{definition}{Definition}
\usepackage{listings}
\usepackage[utf8]{inputenc}
\usepackage{graphicx,stackrel}
\usepackage{dcolumn}
\usepackage{bm}
\usepackage[bbgreekl]{mathbbol}
\usepackage{amsmath,dsfont}
\usepackage[normalem]{ulem}
\usepackage{color}
\usepackage{silence}
\usepackage{float}
\makeatletter
\let\newfloat\newfloat@ltx
\makeatother
\usepackage{algpseudocode}

\usepackage{algorithm}

\algrenewcommand\algorithmicrequire{\textbf{Input:}}
\algrenewcommand\algorithmicensure{\textbf{Output:}}
\usepackage{amsmath,amssymb,amsfonts}
\usepackage[utf8]{inputenc}
\usepackage{subcaption}

\usepackage{hyperref}
\usepackage{multirow}
\usepackage[singlelinecheck=false]{caption}
\usepackage{ragged2e}
\usepackage{ifthen}
\newcounter{complete}
\begin{document}
\title{A quantum pricing-based column generation framework for hard combinatorial problems}

\author{Wesley da Silva Coelho}
\email{wesley.coelho@pasqal.com}
%\thanks{corresponding author}
%\email{wesley.coelho@pasqal.com}
\affiliation{PASQAL SAS, 7 rue Léonard de Vinci, 91300 Massy, France}
\author{Loïc Henriet}
%\email{loic@pasqal.com}
\affiliation{PASQAL SAS, 7 rue Léonard de Vinci, 91300 Massy, France}
\author{Louis-Paul Henry}
\affiliation{PASQAL SAS, 7 rue Léonard de Vinci, 91300 Massy, France}

\date{\today}
\begin{abstract}
In this work, we present a complete hybrid classical-quantum algorithm involving a quantum sampler based on neutral atom platforms. This approach is inspired by classical column generation frameworks developed in the field of Operations Research and shows how quantum procedures can assist classical solvers in addressing hard combinatorial problems. 
We benchmark our method on the Minimum Vertex Coloring problem and show that the proposed hybrid quantum-classical column generation algorithm can yield good solutions in relatively few iterations. We compare our results with state-of-the-art classical and quantum approaches.
\end{abstract}
\maketitle
%--------------------------------------------------------
%--------------------------------------------------------
%--------------------------------------------------------

\section*{Introduction} 

Combinatorial optimization is at the heart of many real-world problems. It consists in finding the ``best" out of a finite, but prohibitively large, set of options.
Column generation~\cite{bibitem_gera1} is an iterative method that was developed to solve this kind of difficult mathematical problems, such as linear formulations where the problem may be too large to consider all options explicitly. 
In this method, variables are associated with each option. The algorithm starts by solving the considered problem with a limited set of variables (or options), known as~\textit{restricted master problem} (RMP), and then iteratively adds variables to improve the objective function. 
The generation of new variables is done by an algorithm specifically tailored to this task: during each iteration, the related new sub-problem to be solved, usually referred to as \textit{pricing sub-problem} (PSP), relies on the duality theory~\cite{bachem1992linear} to provide new variables, if there exists any, only if they can improve the current solution of the restricted master problem. 
The iterative process stops when new variables cease to improve the objective function, which is proven mathematically. 
However, solving the pricing sub-problems usually represents the bottleneck of the column generation (CG) approach as it comes to solving several simpler, but still hard, optimization problems. Hence, designing an efficient way to solve the pricing sub-problems is the most important step to ensure high-quality solutions to the RMP while minimizing time and resource consumption.

During the past years, both academic and industrial communities have been putting a lot of effort into designing quantum hardware and algorithms that could provide a real advantage over classical computers. As pointed out in \cite{pasqore}, this advantage can take the form of more accurate results, a faster convergence, or even a lower energy consumption. Such quantum algorithms can be used along with state-of-the-art classical solutions, such as the column generation algorithm, to create powerful hybrid classical-quantum frameworks.
A wide spectrum of quantum computing platforms is currently being developed, using different kinds of two-level systems as qubits, including Josephson junctions~\cite{Kjaergaard2020, Devoret2004}, trapped ions~\cite{PRXQuantum.2.020343, Benhelm_2008}, photons~\cite{Anton:19}, or neutral atoms~\cite{henriet2020quantum, graph_rydberg1}.
Each of these allows for different quantum processing unit (QPUs) architectures, with their own advantages and limitations when it comes to the connectivity of the qubits, or the types of operations that are easily implemented.
A good knowledge of these platforms allows for the development of hardware-efficient approaches, designed specifically for each of them.
In particular, this allows for identifying the classical bottlenecks that are best suited for being replaced by a quantum approach.

 %In this work, we propose a hybrid algorithm based on neutral atoms QPUs. This platform is particularly well suited to the study of graph optimization problems.

%\subsection*{Our contributions}
\begin{figure}[t]
\centering
\includegraphics[width=0.48\textwidth]{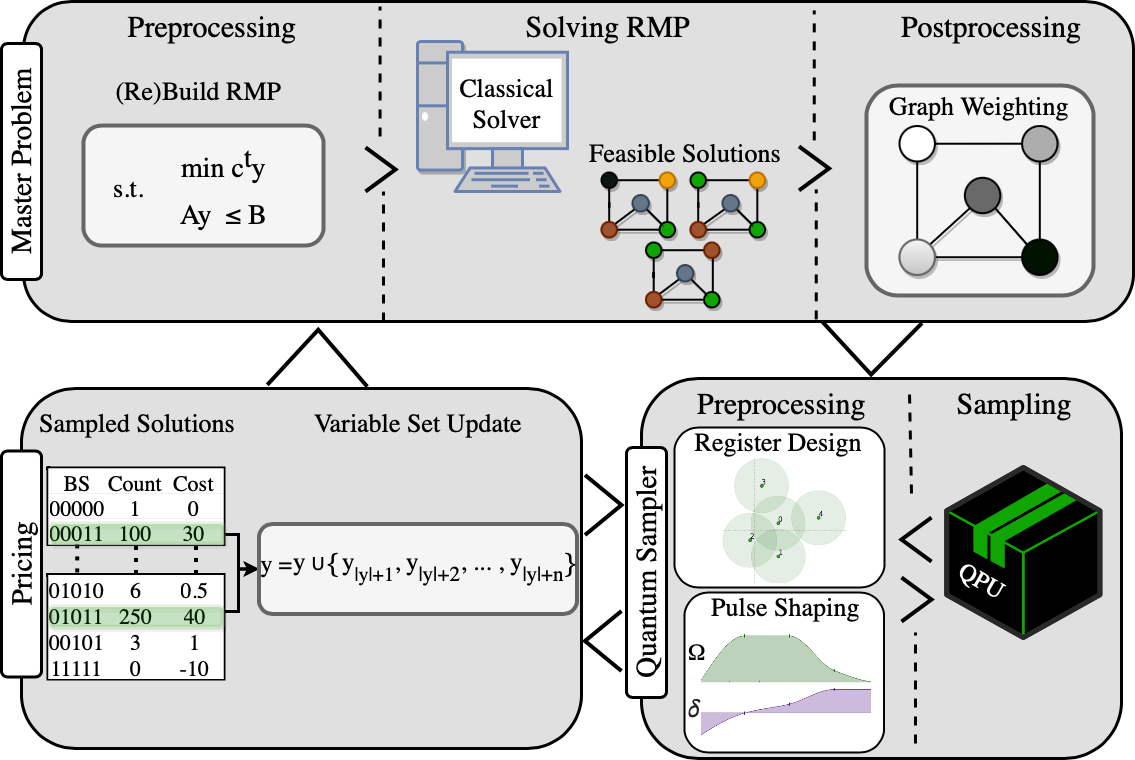}\captionsetup{justification=Justified}
\caption{Workflow of the hybrid classical-quantum column generation approach. First, a minimal sub-set $y$ of variables is generated in such a way that it ensures a feasible solution for the Reduced Master Problem (\textit{e.g.}, with only the singletons of the graph). The RMP is then solved by a classical solver.  The next steps are related to the pricing sub-problems, which are solved by considering the dual values from the solved RMP in order to find more variables that can potentially improve the current solution of the RMP. If such variables exist, then  RMP is updated with the new variables and is solved again. The search for new variables is done by a quantum sampler specifically tailored to consider different inputs related to each pricing iteration. These last steps are repeated until no column is generated by the PSP. }\label{workflowintro}
\end{figure}

In this work, we propose a complete hybrid classical-quantum column generation framework whose pricing sub-problems can be efficiently solved on neutral atom-based QPUs. Unlike other hybrid approaches like QAOA, the core part of the resolution is here carried out by a classical solver, and the quantum processing unit is used as a sampler to restrict the search space. Requiring only $|V|$ qubits for a given graph~$\Graph~=~(\vertices,\edges)$, the related pricing sub-problems are then solved by a neutral atom-based sampler specifically tailored to improve the current solution of the associated master problem. Compared to classical and quantum greedy approaches, we show numerically that the proposed hybrid column generation can improve significantly the quality of the solutions while reducing the number of iterations on the quantum device. Finally, by taking advantage of some quantum features (\textit{e.g.}, state superposition), we find that the hybrid column generation method returns the (near-)optimal solution faster than the classical one (\textit{i.e.}, where no QPU is involved). Fig.~\ref{workflowintro} summarizes our proposed approach.

This paper is structured as follows : 
We first introduce the main aspects of the graph theory and the related combinatorial problems in Section~\ref{background}. After reviewing related works in Section~\ref{soat}, we give a brief introduction to neutral atom-based quantum computing the Section~\ref{rydberg}. %While a greedy algorithm is presented in Section~\ref{greedyyy}, 
We dedicate Section~\ref{methods} to introduce the main idea of the column generation approach. In Section~\ref{solving}, we present in-depth our proposed hybrid classical-quantum approach to solving the Minimum Vertex Coloring Problem (MVCP), while the results of our numerical experiments are discussed in Section~\ref{results}.

\section{Background} \label{background}
Combinatorial problems~\cite{bondy1976graph} have been extensively studied by both academic and industrial communities, and have a vast range of applications in real-world systems. Those problems can naturally be defined on graphs, which are data structures composed of a set of elements called \textit{vertices} (also known as nodes) that can potentially be connected. These connections are called \textit{edges} and might potentially encode different information, such as the importance of such connections (as weights) or the distance between their endpoints. Similarly, different labels and weights can also be associated with vertices in order to differentiate them. Formally, a graph $\Graph = (\vertices,\edges)$ is composed of a set of vertices $\vertices$ and edges $\edges\in\vertices^2$ representing the existence of a connection between vertices $u$ and $v$ from $\vertices$.

Several real-world optimization problems, from a vast spectrum of fields, can be mapped to graph problems. For instance, graphs can be used to encode social experiments~\cite{roth2010suggesting}, telecommunication networks~\cite{9625601}, and physical systems~\cite{barahona1988application}.
The related optimization problems typically consist in selecting a subset of vertices and/or edges optimally satisfying certain rules.
This kind of discrete optimization problem is highly relevant for Quantum Computing (QC), particularly in the case of Noisy Intermediate-Scale Quantum (NISQ)-era platforms~\cite{Bharti_2022, Nishi_2021}.
In that case, results are typically obtained via repeated measurements of the final state of the system. The solutions are then inferred by the selection of the {\it best} sampled state through computationally cheap classical post-processing.

In the case of neutral atom QPUs, the spatial arrangement of qubits can be made such that the Ising Hamiltonian describing the interactions in the system is closely related to a given cost function to be minimized. This is what makes this platform notably well suited to solving graph combinatorial problems~\cite{graph_rydberg1, graph_rydberg2, graph_rydberg3,qeqq}.
As the state of the computational basis in which the qubits are measured has a direct correspondence to the solution to the graph problem, this type of QC is particularly robust to noise (noise can even be an advantage~\cite{Novo_2018}).
For instance, Maximum Independent Set~\cite{dalyac2021qualifying} and Maximum Cut~\cite{pasqore}  problems can be efficiently solved by approaching the ground state of the quantum system with adiabatic annealing~\cite{graph_rydberg2,QA1} or similar methods.
In the following, we formally defined some graph problems and show how they can be solved by quantum-based approaches.

\subsection{Combinatorial problems}
In the following, we present the Maximum Independent set, which is a fundamental part of the proof of concept of our hybrid approach. Then, we formally define the Vertex coloring problem and present the related mathematical formulation.

\subsubsection{Maximum Independent Set problem}
An independent set in a graph $\Graph = (\vertices, \edges)$ is a subset of vertices $\tilde\vertices \subset \vertices$ such that no pair of elements from $\tilde\vertices$ is connected by an edge. The independent sets of $\Graph$ can formally be defined as follows:
 \begin{equation}\label{ISdef}
 IS_{\Graph} = \left\{\tilde\vertices\subset\vertices \bigm| \ \tilde\vertices^2\cap\edges = \emptyset\right\}
 \end{equation}
where $\tilde \vertices^2$ are all the possible edges connecting the vertices in $\tilde \vertices$. Therefore, the Maximum Independent Set~(MIS) is the largest set of $IS_G$:
 \begin{equation}
 \text{MIS}(\Graph) = \underset{\tilde\vertices\in IS_G}{\argmax} \;|\tilde\vertices|.
 \end{equation}
 
The MIS problems can be alternatively described in terms of their quadratic unconstrained binary optimization (QUBO) formulations. 
Consider a graph $\Graph=(\vertices,\edges)$. Let $x_u$ be a binary variable associated with each vertex $u \in \vertices$, and that holds 1 if vertex $u$ is selected to be in the independent set, and 0 otherwise. Hence, the binary vector $\mathbf{x} = \{ x_1, \ldots, x_{|\vertices|}\}$ can be put in one-to-one correspondence with partitions $\tilde \vertices$ of the vertex set $\vertices$ via the identification:
\begin{equation}
 \tilde \vertices (\mathbf{x}) = \left\{u \in \vertices \bigm| \ x_u = 1 \right\}
\end{equation}
The solution to the maximum (weighted) independent set problem is then given by:
\begin{equation}\label{eq:misqubo}
\hspace{-0.3cm} \text{MIS}(\Graph) = \underset{\mathbf{x} \in \{0,1\}^{|\vertices|}}{\argmin} \left( -\sum_{u \in \vertices} w_ux_u + \alpha\sum_{\{u,v\} \in \edges} x_u x_v \right)
\end{equation}

where the parameter $w_u$ represents the weight associated to each vertex $u\in \vertices$, while $\alpha > 0$ is an arbitrary coefficient to penalize unfeasible solutions. %Note that, on unweighted graphs, the penalty coefficient $\alpha$ as well as all $w$ parameters are set to 1.
Note that, on unweighted graphs, all $w$ parameters are set to 1, while the penalty coefficient $\alpha$ must hold any value equal to or greater than 2.

\subsubsection{Minimum Vertex Coloring problem}

 The Vertex Coloring problem has several applications in real-world optimization problems such as network design~\cite{cornejo2010deploying} and task scheduling~\cite{marx2004graph}. A vertex coloring is an assignment of colors (or labels) to each vertex of a graph such that any two identically colored vertices are not connected by an edge. The \textit{Minimum Vertex Coloring} problem consists then in finding a feasible coloring while minimizing the number of colors (or labels) assigned; the minimum number of colors used to color all vertices of a given graph $\Graph$ is called its chromatic number, hereafter denoted $\mathcal{X}(\Graph)$. The Minimum vertex coloring can be formally defined as follows:

\begin{definition}
Let $\Graph =(\vertices,\edges)$ be a graph with a set $V~=~\{u_1, .. , u_{|\vertices|}\}$ of vertices and a set $\edges \subset \vertices^2$ of edges. Also, let $C$ be a set of available colors. The Minimum Vertex Coloring Problem consists in coloring each vertex of $\Graph$ with exactly one color from $C$ in a such way that the number of used colors is minimized while ensuring that no two adjacent vertices have the same color.
\end{definition}

\begin{figure}[t] \centering
 \begin{subfigure}[b]{0.3\linewidth}
 \centering
 \captionsetup{width=1.15\linewidth}
 \includegraphics[width=\linewidth]{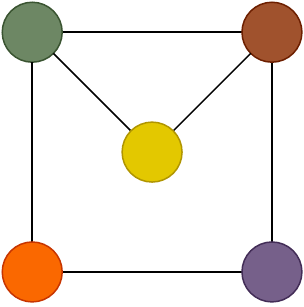} 
 \caption{}
 \label{fig:trivial}
 \end{subfigure}\hspace{+2cm}
 \begin{subfigure}[b]{0.3\linewidth}
 \centering 
 \captionsetup{width=1.15\linewidth}
 \includegraphics[width=\linewidth]{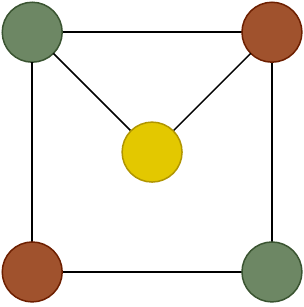} 
 \caption{}
 \label{fig:opt}
 \end{subfigure}
 \captionsetup{justification=Justified}
 \caption{Two coloring solutions to the same graph with 5 vertices, 6 edges, and $\mathcal{X}(\Graph) = 3$. The set $C$ has 5 available colors: green, brown, orange, purple, and yellow. Figure \ref{fig:trivial} shows a trivial coloring, where each color is mapped to exactly one vertex, while Figure \ref{fig:opt} depicts an optimal solution for the same instance.}
 \label{fig:ex}
\end{figure}

Figure~\ref{fig:ex} shows two possible coloring solutions for the same graph. By applying a trivial coloring (see Fig.~\ref{fig:trivial}), each color is mapped to exactly one vertex; this simple approach always gives a feasible solution to the problem. As shown in Fig.~\ref{fig:opt}, however, the related chromatic number (i.e., the optimal solution) can be reduced to 3.
It is worthwhile to notice that, given a feasible solution for the Vertex Coloring problem, any sub-set of vertices colored with the same color is also an independent set. 
Hence, finding the minimum sub-set of independent sets that cover all vertices of a given graph~$\Graph$ is equivalent to solving the MVCP in the same graph. Finding the chromatic number of a graph, however, is one of Karp's 21 NP-complete problems~\cite{karp1972reducibility}.

\subsection{An extended formulation for the Minimum Vertex Coloring problem}
We now present an extended formulation\footnote{Extended formulations are mathematical models in which the number of variables grows exponentially as the input increases.} for the Minimum Vertex Coloring Problem, which is used within our proposed hybrid approach. First, let $S$ be a set of all possible independent sets in the graph $\Graph = (\vertices,\edges)$. Also, let $b_{us}$ be a binary parameter that holds 1 if the vertex $u \in V$ is present in the independent set $s \in S$, and 0 otherwise. Finally, we associate a binary variable~$y_s$  to each independent set $s \in S$; it takes~1 if the related independent set is selected; 0 otherwise. Solving the MVCP comes then to solving the following extended formulation:
 
\begin{align}
 & min \sum_{s \in S}y_s \label{foe}
\end{align}
s.t.,
\begin{align}
 &\sum_{s \in S }b_{us}y_s = 1, &\forall u \in \vertices\label{ce} \\
 &y_s \in \{0,1\} , &\forall s \in S\label{cei}
\end{align}

where~\eqref{foe} is set to minimize the number of selected independent sets while ensuring that each vertex of the graph is present in exactly one of them (see equation~\eqref{cei}).  Note that, by considering each independent set $s \in S$ as a color assignment, the adjacency constraints related to the Minimum Vertex Coloring problem are automatically respected (see definition \eqref{ISdef}). 

 The number of all independent sets in a graph, and hence the number of $y_s$ variables, can be extremely large, exponentially growing as the number of vertices in the graph increases. Hence, as one may anticipate, finding all such sets on a given graph is a very hard task and can be very time and resource-consuming even for small instances. To overcome the aforementioned limitations, we propose a hybrid classical-quantum column generation-based framework to efficiently solve the proposed extended formulation by enumerating only a small subset of independent sets. In what follows, we present the related works and discuss how a quantum sampler can be integrated into classical frameworks.   

\section{Related work}
\label{soat}
Several quantum algorithms have been proposed for solving graph coloring problems in the past years, and most of them rely on a quantum annealing-based approach. In~\cite{kudo2018constrained}, the authors investigate a real-time quantum dynamics-based quantum annealing approach where the related Hamiltonian is designed to naturally respect all problem-related constraints without adding penalty terms. Authors in~\cite{ardelean2022graph}, on the other hand, propose a genetic algorithm-based quantum approach to solve both vertex and edge coloring problems in different highly configurable circuit-based models. Titiloye and Crispin~\cite{titiloye2011quantum} compare classical and quantum annealing approaches in solving graph coloring problems. According to the authors, the path-integral Monte Carlo-based quantum annealing (QA) algorithm outperforms its classical counterpart. 

Authors in~\cite{silva2020mapping} propose another approach in which the problem-related set of constraints is transformed into an energy minimization problem to output a QUBO formulation, which is then solved in a quantum annealer platform. However, by running various numerical simulations and comparing results obtained with standard and enhanced circuit-based QAOA algorithms, authors in~\cite{tabi2020quantum} indicate the limitation of the existing QA hardware solutions for solving the Vertex Coloring problem. Also, Silva et al~\cite{kk} compare simulated and quantum annealing approaches for solving the proposed QUBO formulation for the Vertex Coloring problem. Using D-Wave 2X as an independent set sampler for a simple greedy framework, the authors show that the proposed quantum sampler could improve the results with high probability on small graphs due to hardware limitations.

Fabrikant and Hogg~\cite{fabrikant2002graph} introduce a quantum heuristic for graph coloring for instances that can be solved with at most 3 colors. Using two qubits to each vertex of the graph, an approximate asymptotic analysis suggests polynomial-time cost for solving the related 3-coloring problem. Moreover, authors in~\cite{shimizu2022exponential} introduce an exponential-space quantum algorithm to solve the MVCP in $O(1.9140^{|\vertices|})$ running time. They propose a quantum random access memory framework based on Ambainis quantum dynamic programming~\cite{ambainis2019quantum} with applications of Grover’s search to branching algorithms. Moreover, authors in~\cite{21:sc:quantum} proposed a greedy quantum algorithm for solving the MVCP by iteratively computing the solutions of Maximum Independent Set problems. By simulating a framework on a classical computer to reproduce the Rydberg blockade phenomenon on neutral atoms-based QPUs, they show that their approach can always find feasible solutions for the problem. More details about their approach are presented in Appendix~\ref{greedyyy}. 

Finally, authors in~\cite{ossorio2022optimization} propose a quantum annealing-based method to solve one of the two pricing sub-problems of a column generation-based approach for the refinery scheduling problem. They first decompose the problem into one master problem and two pricing sub-problems and formulate one of them as a QUBO model. While only the QUBO-related pricing sub-problem is solved using a quantum annealing approach, the master and the other pricing sub-problem are solved with a classical solver. The quantum system is based on D-Wave's quantum annealers and is designed to return only the optimal solution. Due to some serious limitations either related to the problem size or to the connectivity of its variables, the authors could guarantee high-quality solutions for only a few instances of small graphs.

Note that, even though those works propose interesting approaches to solving combinatorial problems (sometimes only the decision version), little attention has been given to hybrid and analog approaches, especially using neutral atoms-based QPUs. In what follows, we introduce a complete quantum pricing framework based on neutral atom QPUs that can be easily embedded into a column generation algorithm to solve hard combinatorial problems, such as graph coloring problems.

\section{Neutral atom QPUs} \label{rydberg}

In neutral atom-based QPUs, lasers or microwaves are used to induce transitions between electronic states of the valence electron of alkali metal (typically Rubidium) atoms.
Different pairs of electronic levels can be used as qubits. 
Here, we will solely focus on the case where those two states are the electronic ground state $\ket{g}\equiv\ket{0}$ and a $s-$ Rydberg level $\ket{r}\equiv\ket{1}$.
In that case, atoms can be placed arbitrarily in space, so that the effective Hamiltonian of the atoms at time $t$ can be written as
\begin{equation}\label{eq:hamiltonian}
 H(t) = \Omega(t)\sum_{u=1}^{|\vertices|}\hat{\sigma}^x_u - \Delta(t)\sum_{u=1}^{|\vertices|}\hat{n}_u +\sum_{u<v=1}^{|\vertices|} U_{uv} \hat{n}_u \hat{n}_v,
\end{equation}
where the amplitude (giving the Rabi frequency) $\Omega(t)$ and detuning $\Delta(t)$ of the laser can be controlled over time, and the interaction strength 
$U_{uv}\propto\left|{\bf r}_u-{\bf r}_v\right|^{-6}$ is a function of the distance between atom~$u$ and atom~$v$. Throughout this paper, we set $\hbar=1$.
Note that in the current work, we consider only a uniform global laser control over the atoms.

A key property of Rydberg physics is the so-called Rydberg blockade mechanism~\cite{henriet2020quantum} : 
the two-body interaction term in (\ref{eq:hamiltonian}) forbids the simultaneous excitation of two atoms that are closer than a certain distance.
Given a set of atoms and their positions, one can then define a graph such that each atom corresponds to a vertex and in such a way that two vertices are connected by an edge if and only if the distance between the related atoms is shorter than a given threshold.
This kind of graphs are known as Unit-Disk graphs, and they are the most natural graphs to encode in a neutral atom QPU.
For this graph class, one can ensure that the evolution of the quantum system is restricted to a subspace of the complete Hilbert space corresponding to independent sets of the graph. By setting $\Omega(t)=0$ and adjusting the value of $\Delta$, one can ensure that the ground-state corresponds to an MIS.
Because of these properties, people have explored Quantum Annealing as a way to solve optimization problems on graphs~\cite{Pichler}. For non-UD graphs, however, one can construct alternative approaches, similar to what is done in QAOA, or Variational Quantum Eigensolvers~\cite{qscore}.

\section{Problem decomposition}
 \label{methods}

We dedicate this section to fully describing the decomposition of the proposed extended formulation~\eqref{foe}-\eqref{cei}, a fundamental step to solve the related combinatorial problem with a column generation-based algorithm. The need of applying such a mathematical strategy comes from the fact that, in most of cases, generating all elements that will be related to the variables of an extended formulation is a very hard task. For instance, enumerating all independent sets of a graph can be impractical even for small instances. To overcome the aforementioned issue, we decompose the problem under consideration into two problems, named \textit{Restricted Master Problem} (RMP) and \textit{Princing Sub-Problem} (PSP). While the former has only a small subset of variables needed to find a solution to the problem, the latter is designed to provide new elements (e.g., independent sets) that respect all technical constraints imposed by the RMP. These new elements are then added as new variables (also seen as columns) to the mathematical model related to the RMP as they might potentially improve the quality of the solution (e.g., decreasing the number of colors needed to solve the Vertex Coloring problem).

Column generation-based approaches rely on the duality theory~\cite{balinski1969duality}, which states that optimization problems can be addressed from two different perspectives: the \textit{primal problem} or its counterpart, the \textit{dual problem}. The relationship between these two problems is the following: (i) for each variable (resp.~constraint) in the primal problem, there is a related constraint (resp.~variable) in the dual problem, and (ii) the optimization direction (\textit{e.g.}, maximization or minimization) on the dual is inversed related to its primal counterpart. For each sub-optimal solution that satisfies all the constraints on the primal problem, there is at least one direction to move in such a way that the objective function is improved. Such improving directions are represented by the vector of dual variables and optimizing them is equivalent to tightening the bounds of the primal problem. For an in-depth discussion on the primal-dual relationship, one may refer to~\cite{balinski1969duality}. After solving a linear model related to a primal problem, one can easily get such a direction vector (\textit{i.e.}, dual variables) by calling some built-in function proposed by the solver used in the process.

In order to design an efficient column generation framework, the PSP is then formulated in such a way to incorporate the dual information provided by current solutions of the RMP, which implies solving a different pricing instance each time the sub-problem is called. Hence, by applying the duality theory, the PSP searches for new variables that can improve the objective function of the RMP, and once it is mathematically proven that it is no longer possible to generate such variables (\textit{i.e.}, new columns), the loop-based procedure stops. This approach is very powerful when only a few variables are normally activated (\textit{i.e.}, taking any value other than zero) in the optimal solution to a given combinatorial problem.  Hence, applying such a technique, only a very small subset of variables is needed to be generated by the pricing routine. 

%\subsection{Problem decomposition}
In what follows, we present a decomposition scheme for solving the proposed extended formulation for the MVCP. The main idea relies on generating a restricted model with only a sub-set of independent sets and iteratively updating it with new variables (\textit{i.e.}, columns) that have the potential to improve the current solution. 

\subsection{The restricted master problem}

Since $\Graph$ has an exponential number of potentially suitable colorings represented by the related set $S$ of independent sets, the extended formulation~\eqref{foe}-\eqref{cei} admits an exponential number of variables. To overcome the difficulty of generating $S$, we propose to generate only a small subset $S' \subseteq S$ of variables that are needed to solve the master problem. For instance, a trivial solution might be initializing $S'$ with only the singletons of the input graph. The reduced model is then hereafter referred to as \textit{Restricted Master Problem} (RMP), and can be defined as follows:

\begin{align}
 & min \sum_{s \in S'}y_s \label{rfoe}
\end{align}
s.t.,
\begin{align}
 &\sum_{s \in S' }b_{us}y_s = 1, &\forall u \in \vertices\label{rce1} \\
 & 0 \leq y_s \leq 1 , &\forall s \in S'\label{rce}
\end{align}

Note that, in order to apply the duality theory, we must solve the linear relaxation on the RMP, meaning the $y$ variables are no longer binary in this formulation. The RMP is then solved again with the integrality constraints \eqref{cei} once it has all variables needed to provide the optimal solution for the relaxed RPM, which can be proven mathematically; we provide this proof in the following section.

\subsection{Pricing sub-problems}
We present first the dual formulation related to the relaxed RMP~\eqref{rfoe}-\eqref{rce}. By associating a dual variable $w_u$ to each constraint in~\eqref{rce1}, we define the dual problem as follows:

\begin{align}
 & max \sum_{u \in \vertices}w_u \label{dfoe}
\end{align}
s.t.,
\begin{align}
 &\sum_{u \in \vertices}w_ub_{us} \leq 1, &\forall s \in S'\label{dce} \\
 &w_u \in \mathcal{R}, &\forall u \in \vertices\label{dcei}
\end{align}

The separation of inequalities~(\ref{dce}) represents the pricing sub-problems related to the extended formulation~\eqref{rfoe}-\eqref{rce}. The PSP consists then in finding a new coloring set in such a way that all adjacency constraints would be respected while improving the solution cost of RMP (\textit{i.e.}, decreasing the value of the objective function~(\ref{rfoe})). For this purpose, let $\bar{w}_{u}$ be the components of the current dual solution of the RMP related to constraints~(\ref{rce1}).  By setting each dual variable~$\bar{w}_u$ as the weight of the related vertex $u \in \vertices$, finding a new independent set under such conditions comes to solving the following maximum weighted independent set (MWIS) formulation: \vspace{-0.5cm}

\begin{align}
 & max \sum_{u \in \vertices}\bar{w}_{u}x_u \label{pfo}
\end{align}
s.t.,
\begin{align}
 & x_u + x_v \leq 1, & \forall (u,v) \in  \edges\label{pc}\\
 &x_u \in \{0,1\}, & \forall u \in \vertices\label{pv}
\end{align}
where $x_u$ is a binary variable that holds 1 if it is in the independent set; 0 otherwise. Inequalities~\eqref{pc} ensures that the new independent set respects the adjacency constraints. Note that the pricing formulation~\eqref{pfo}-\eqref{pv} is an Integer Programming (IP) version of the QUBO formulation~\eqref{eq:misqubo}.

The net gain of adding a new variable related to a solution to the pricing problem is given by the reduced cost. Based on the separation of inequalities~(\ref{dce}), we calculate the reduced cost $r_s$ for any independent set $s$ given by solving the formulation \eqref{pfo}-\eqref{pv} as following:

\begin{align}
    & r_s  = 1 - \sum_{u \in \vertices}\bar{w}_{u}x_u \label{red_cos}
\end{align}

 Since we minimize the master problem in the work, any solution with a negative reduced cost might potentially improve the solution of the RMP. The pricing problem consists then in finding a new independent set~$s$ whose total weight is strictly greater than 1; the total weight of any independent set is calculated as in the cost function \eqref{pfo}. Hence, any solution to the formulation~\eqref{pfo}-\eqref{pv} whose cost is greater than 1 can therefore be added to the sub-set $S'$; if such a solution does not exist, then the solution of the RMP cannot be improved and, hence, the optimal solution the relaxed RPM can be achieved with the current variables related to the sub-set $S'$.

\section{Solving method}
\label{solving}

\begin{figure}[t]
\begin{center}
\includegraphics[scale = 0.6]{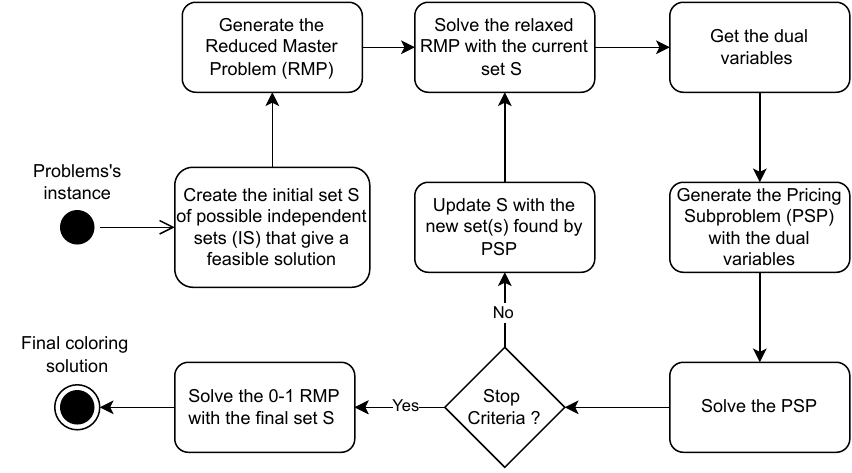}
\captionsetup{justification=Justified}\caption[Interaction between the restricted master problem and the pricing sub-problem]{Interaction between the restricted master problem and the sub-problem}
\label{ipmsp}
\end{center}
\end{figure}

As previously discussed, solving the pricing sub-problems is usually the bottleneck in column generation-based algorithms since it comes to solving different instances of a hard combinatorial problem multiple times. To overcome this problem, we propose a quantum pricing algorithm that can find the (near-) optimal solution faster than the classical one (\textit{i.e.}, where no QPU is involved). For this purpose, let us now describe the column generation-based framework proposed to solve the Minimum Vertex Coloring problem; which is summarized in Fig.~\ref{ipmsp}. 

First, a minimal sub-set~$S' \subseteq S$ of independent sets is generated in such a way that it ensures a feasible solution for the extended formulation~\eqref{foe}-\eqref{cei}. As previously discussed, the most trivial way to build the initial set~$S'$ of independent sets is generating only the singletons in the graph; this simple approach always provides a solution for the RMP.

The classical part of the proposed hybrid approach is related to the Restricted Master. Once the initial set~$S'$ is created, the RMP is built and then solved on its linear relaxation form (see formulation~\eqref{rfoe}-\eqref{rce}) by a classical solver (\textit{e.g.}, GPLK). The values of the dual variables are also given by the classical solver by running a built-routine after solving each version of the RMP (i.e., with different sub-sets of variables).

The next steps are related to the pricing sub-problems, in which the PSP is solved by applying the values of the related dual variables from the solved RMP. As previously discussed, this step comes to finding independent sets whose weight is strictly greater than 1. If such elements exist, then they are added to~$S'$. As we detail in the following, we propose a quantum sampler that is specifically tailored to output multiple independent under the aforementioned conditions For each new independent set found by solving the related pricing sub-problem, a new variable is created and added to the sub-set~$S'$. Then, the RMP is solved again with the new columns (\textit{i.e.}, independent sets converted into variables). These last steps are repeated until no column is generated by the PSP. Finally, the final RMP is solved with all generated variables (\textit{i.e.} independent sets) with the integrality constraints~\eqref{cei}, as previously discussed.

\subsection{Worked example}\label{wordewclassic}
\begin{figure}[b] 
 \centering
 \includegraphics[width=0.35\linewidth]{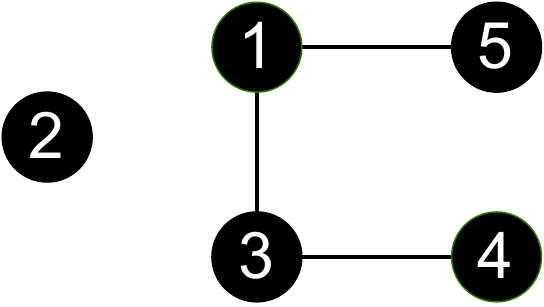} 
 \captionsetup{justification=Justified}
 \caption{Illustration of a graph with 5 vertices and 3 edges; the set~$S$ of all independent sets is composed by the five singletons and the sub-sets [1,2], [1,4], [2,3], [2,4], [2,5], [3,5], [4,5], [1,2,4], [2,3,5], [2,4,5].  }
 \label{workex} 
\end{figure}

 Table \ref{tab:my-table} shows a worked example of applying the column generation framework on the graph represented in the Fig.~\ref{workex}, where the RMP is solved classically by an IP solver. The first column shows how many PSPs were solved before reaching the final solution. The second column depicts the independent sets selected as the solution for the RMP formulation~\eqref{rfoe}-\eqref{rce}. The third column presents the dual solution given by the variables~$w$ (see the dual formulation~\eqref{dfoe}-\eqref{dcei}). Finally, the last column depicts the maximum weighted independent sets generated after running each PSP, where the~$\bar w$ parameters in the cost function of the formulation~\eqref{pfo}-\eqref{pv} are set to the values of dual variables~$w$; the generated weighted independent set is then added to the set~$S'$ before rerunning the RMP. The final coloring is represented by the last solution for the RMP, where one color is given for each selected independent set.
 
In this example, we generate all five singletons of the graph as the first sub-set~$S'$ of independent sets before solving the first version of the RMP. As observed, only 4 more independent sets (out of the remaining 10 to be generated) were needed to find the best solution. Indeed, even though the closed-loop could be stopped after iteration 4, the optimality was only proven after the fifth iteration, when no independent set whose total weight is greater than 1 can be generated (see the dual solution on the last row, which is used as vertex weights). Note that the independent set generated in the i-th iteration is likely to be selected in the solutions of the RPM in the following iteration.

The values of the dual variables can be calculated by solving the dual formulation~\eqref{dfoe}-\eqref{dcei}, where the cost function is set to be equal to the solution cost of the RMP in the same interaction. In this example, the cost function of the dual formulation was set to be equal to 5, 3, 3, 2, and 2 in the first, second, third, fourth, and fifth iterations. However, most commercial solvers solve the dual formulation in parallel in order to prove the optimality of the (RMP) primal solution. Hence, the final value of the dual variables can then be easily provided by a built-in solver's function (e.g., \textit{solver.get$\_$dual()}). Note, however, that only one independent set (the most weighted one) is generated by solving the related pricing formulation~\eqref{pfo}-\eqref{pv}  and, due deterministic nature of this approach, the final solution is always the same if the input (\textit{i.e.}, the graph and vertex weights) does not~change. 

\begin{table}[b]
\resizebox{\columnwidth}{!}{%
\begin{tabular}{cc|c|c}
\textbf{Iteration} & \textbf{RMP solution} & \textbf{Dual solution} & \textbf{MWIS} \\ \hline
\multicolumn{1}{c|}{1} & [1] , [2] , [3] , [4] , [5]   & [1.0, 1.0, 1.0, 1.0, 1.0] & [1, 2, 4] \\
\multicolumn{1}{c|}{2} & [3] , [5] , [1, 2, 4]  & [1.0, 1.0, 1.0, -1.0, 1.0] & [2, 3, 5] \\
\multicolumn{1}{c|}{3} & [3] , [5] , [1, 2, 4] & [1.0, -1.0, 1.0, 1.0, 1.0] & [1, 4] \\
\multicolumn{1}{c|}{4} & [1, 4] , [2, 3, 5] & [1.0, -1.0, 1.0, 0.0, 1.0] & [3, 5] \\
\multicolumn{1}{c|}{5} & [3, 5], [1, 2, 4] & [1.0, 0.0, 1.0, 0.0, 0.0] & None
\end{tabular}%
}
\captionsetup{justification=Justified}\caption{A worked example of applying the column generation framework on the graph represented in the Fig.~\ref{workex}. Only 5 independent sets (out of 10) are generated to find the optimal solution.}
\label{tab:my-table}
\end{table}

% \usepackage{graphicx}
\begin{table}[t]
\resizebox{\columnwidth}{!}{
\begin{tabular}{cc|c|c}
\textbf{Iteration} & \textbf{RMP solution} & \textbf{Dual solution} & \textbf{IS generated} \\ \hline
\multicolumn{1}{c|}{1} & [1],[2],[3],[4],[5] & [1.0, 1.0, 1.0, 1.0, 1.0] & [1, 2, 4],[2, 4, 5] \\\multicolumn{1}{c|}{} &&& [2, 3, 5] , [1, 2]\\ \multicolumn{1}{c|}{} &&&[2, 5] , [2, 3]\\\multicolumn{1}{c|}{} &&&[4, 5] , [3, 5]\\ 
\multicolumn{1}{c|}{2} & [3, 5], [1, 2, 4] & [1.0, -0.5, 1.0, 0.5, 0.0] & [1, 4] \\
\multicolumn{1}{c|}{3} & [1, 4],[2, 3, 5] & [1.0, 0.0, 1.0, 0.0, 0.0] & None
\end{tabular}
}
\captionsetup{justification=Justified}\caption{A worked example of applying a quantum sampler to solve the pricing sub-problems related to the column generation framework. Considering the graph represented in the Fig.~\ref{workex}, only 3 different pricing sub-problems were solved before reaching the final solution. }
\label{tab:my-table2}
\end{table}

Now, let us exemplify how a quantum sampler could be applied to solve the pricing sub-problems within the column generation framework; Table \ref{tab:my-table2} shows such a worked example. As presented in the previous worked example depicted in Table \ref{tab:my-table}, we consider the graph depicted in Fig.~\ref{workex} initial sub-set~$S'$ composed only of singletons. As observed, only three iterations are needed to find the final solution. Indeed, sampling more independent sets on each pricing iteration can considerably improve the performance of the column generation algorithm. In this example, 9 out of 10 possible independent sets were generated. In fact, due to its stochastic nature, a quantum sampler can efficiently provide multiple sets with the same input and, hence, speed up the convergence of the RMP to the optimal solution.

In what follows, we present how neutral atom-based systems can be designed to take into consideration the dual values from the RMP to output multiple weighted independent sets and efficiently solve the related PSPs.

\subsection{Embedding strategy}\label{spring_section}
\begin{algorithm}[b]\footnotesize
\caption{Vertex-atom remapping}\label{alg:permu}
\begin{algorithmic}[1]
\Require A graph $\Graph'$, a register $R$ and vertex weights $W$.
 \Ensure The reduced register $R$.
\State Let $\Bar \vertices'$ be the list of vertices from $\Graph'$ sorted in descending order by the weight given by $W$ 
\State Remove all vertices whose weights are less than or equal to zero from $\Bar \vertices'$
\State Map vertex $\Bar \vertices'.first$ to the farthest position from the center of $R$ 
\State $\Bar \vertices' \gets \Bar \vertices' \backslash \Bar \vertices'.first$ 
\While{$\Bar \vertices'$ has vertices}
\State Map vertex $\Bar \vertices'.first$ to the farthest position from all vertices already mapped to an atom in $R$
\State $\Bar \vertices' \gets \Bar \vertices' \backslash \Bar \vertices'.first$ 
\EndWhile
\State Remove all atoms not mapped to any vertex from register $R$
\State \textbf{return} register $R$
\end{algorithmic}
\end{algorithm}
Embedding random graphs into UD-disk representations is proven to be an NP-hard problem~\cite{breu1998unit}. Also, not all graphs have such a realization. For instance, it is not possible to find a UD-disk representation for any~$K_{1n}$ star graph with~$n>6$ on a 2-dimension plane. In order to design a near-optimal register to represent any graph given as an input, we use the embedding strategy presented in~\cite{pasqore}, where an algorithm based on force-directed principles is used to embed graphs into planes in such a way that two connected (resp.~disjoint) vertices are placed close to (resp.~far from) each other, with a minimum (resp.~maximum) distance between them (resp.~from the plane's center). See Appendix \ref{FRalgo} for more details.

Once the initial register is created as previously discussed, it has the same number of atoms as the number of vertices on the initial graph~$\Graph$. However, solving the pricing sub-problem within the column generation framework might potentially imply finding one or more independent sets on a sub-graph~$\Graph'$: this sub-graph is generated with only the vertices with positive weight (\textit{i.e.}, the positive dual value provided by the solved restricted master problem).
The light pattern holding the atoms in place is created via a spatial light modulator, and determining the right settings for this device demands lengthy calibrations~\cite{qeqq}.

Therefore, creating a new register by running the Fruchterman-Reingold algorithm with the remaining vertices would be too time-consuming. 
%This is due to the requirement to recalibrate the SLM which will be used due to the potential necessity of recall the right settings needed for :  
%Creating a new register by running the Fruchterman-Reingold algorithm with the remaining vertices might potentially imply re-calibrating the quantum device, which is not ideal due to the related time-consuming setup process.
Instead of just removing the atoms mapped to the vertices that are no longer in the sub-graph~$\Graph'$, we propose here a vertex-atom remapping strategy that takes into consideration the vertices' weights (dual values). 

\begin{figure}[t] 
 \begin{subfigure}[b]{0.498\linewidth}
 \centering
 \includegraphics[width=0.55\linewidth]{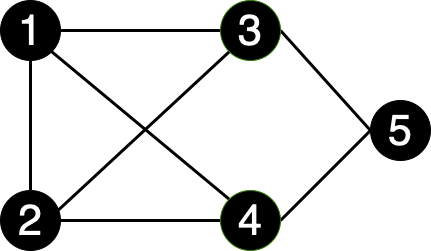} 
 \caption{} 
 \label{gtbe} 
 \end{subfigure}%% 
 \begin{subfigure}[b]{0.498\linewidth}
 \centering
 \includegraphics[width=0.55\linewidth]{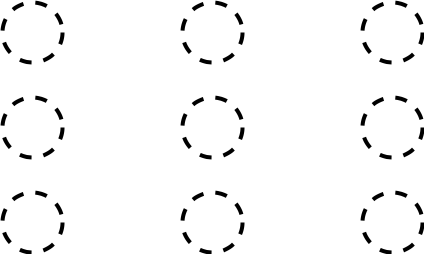} 
 \caption{} 
 \label{initial} 
 \end{subfigure} 
 \begin{subfigure}[b]{0.498\linewidth}\vspace{.51cm}
 \centering
 \includegraphics[width=0.55\linewidth]{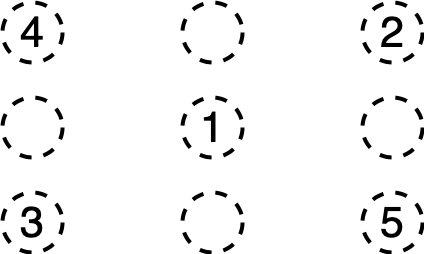} 
 \caption{} 
 \label{map} 
 \end{subfigure}%%
 \begin{subfigure}[b]{0.498\linewidth}
 \centering
 \includegraphics[width=0.55\linewidth]{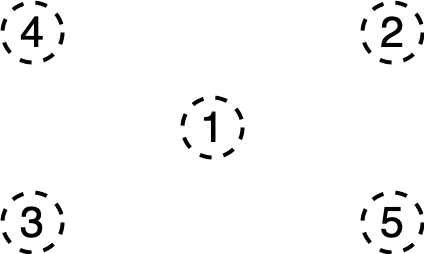} 
 \caption{} 
 \label{final} 
 \end{subfigure} \captionsetup{justification=Justified}
 \caption{Vertex-atom mapping for a 5-vertex graph on a 9-atom register. Fig.~\ref{gtbe} presents a graph with 5 vertices, whose weights are to their indexes (\textit{e.g.}, the weight of vertex 1 (resp.~5) is equal to 1 (resp.~5)). Fig.~\ref{initial} shows a register with 9 atoms in their positions. Fig.~\ref{map} depicts a mapping where the most weighted vertices are far from each other. Finally, Fig.~\ref{final} presents the final register and the vertices from the sub-graph they represent.}
 \label{apa} 
\end{figure}

Algorithm~\ref{alg:permu} summarizes the main idea of our proposed approach: it receives a sub-graph~$\Graph' = (\vertices', \edges')$, a register~$R$ composed of atoms and their positions within the QPU, and vector~$W$ representing the weights for all vertices in~$\vertices'$. First, let~$\Bar \vertices'$ be the list of vertices of~$\Graph'$ sorted in descending order by the weight given by~$W$. Then, remove all vertices whose weights are less than or equal to zero from~$\Bar \vertices'$ (step~2), map the most weighted vertex to the further atom from the register's center (step~3), and remove the remapped vertex from~$\Bar \vertices'$ (step~4). Then, the most weighted vertex in~$\Bar \vertices'$ is embedded into the farthest atom position from all atoms already mapped to a vertex in~$\Graph'$, and removed from~$\Bar \vertices'$; these steps are done until all vertices from~$\Graph'$ are mapped to an atom in the register~$R$. Finally, as seen in~step~9, all atoms not mapped to any vertex are removed from~$R$. Let us recall that, since the proposed remapping strategy is applied to solve the PSPs, the weight vector~$W$ is generated with dual values provided by the solution of the current RMP, as previously mentioned. A worked example of applying the proposed algorithm is depicted in Fig.~\ref{apa}.
 
\subsection{Independent set quantum sampler}\label{quantumsampler}
We propose an independent set sampler based on current-generation neutral atom Quantum Processing Units, which is inspired by the quantum adiabatic approach described by~\cite{aqmis}.
In the latter, the main idea is to slowly evolve the system from an easy-to-prepare ground state to the ground state of the final cost Hamiltonian~$H_C$. 
By slowly evolving the system, the atoms stay in the instantaneous ground state~\cite{albash2018adiabatic}. Here, we only aim at keeping the system {\it close} to the instantaneous ground state, allowing it to pick up components in the low-lying states.

In order to do such evolution, we continuously vary the detuning~$\delta (t)$ and the Rabi frequency~$\Omega( t )$ in time, starting with~$\Omega ( t_0 ) = 0$, $\delta (t_0) < 0$ and ending with~$\Omega( t_f ) = 0 $ , $\delta (t_f) > 0$. Hence, the initial ground-state of~$H ( t_0 )$ is  $| 00000 \bigr \rangle$, while the low-energy space of~$H ( t_f )$ contains that of the cost Hamiltonian~$H_C$.
This protocol originally aims at preparing the ground-state and solving the MIS problem. In this work, we use it to sample independent sets whose total weight is equal to or close to that of the MWIS.

To ensure that we are not exciting the system to states that do not form independent sets, we have to estimate the minimal distance between atoms that are disjoint in the graph (this yields~$\Omega_{d}$), and estimate the farthest distance between two connected atoms, which gives~$\Omega_{c}$. For this purpose, let~$d_{uv}$ be the distance between nodes~$u$ and~$v$ from~$\vertices$, while~$C_6$ is the interaction coefficient related to the quantum device. Then,~$\Omega_{d}$ and~$\Omega_{c}$ are respectively  defined as following:
\begin{align}
& \Omega_{d} = \underset{(u,v) \notin \edges}{\argmax}\quad C_6d_{uv}^{-6} \\
& \Omega_{c} = \underset{(u,v) \in \edges}{\argmin}\quad C_6d_{uv}^{-6}
\end{align}

In unit-disk (UD) graphs, keeping~$\Omega \in [ \Omega_{d} , \Omega_{c} ] $ ensures that only independent sets appear in the dynamics (provided that the dynamics is not too fast)~\cite{Pichler, quantumscars}.
A large value of $\Omega$ is desirable to speed up the dynamics of the system and reach states that are {\it far} from the initial state.
That is why we set~$\Omega_{max} = max(\Omega_{c},\Omega_{d}) $ as the maximum value the Rabi frequency can take during the pulse sequence.

%Regardless of the class of the graph (\textit{i.e.}, whether it is a UD or not), we set~$\Omega_{max} = max(\Omega_{c},\Omega_{d}) $ as the maximum value the Rabi frequency can take during the pulse sequence. 
In the case of non-UD graphs, this value of $\Omega_{max}$ is still a good compromise. Indeed, if $\Omega<\Omega_c$ then the effective graph $\Graph'$ that is encoded in the system contains more edges than the target graph $\Graph$. The independent sets of $\Graph'$ are then not strictly included in the independent sets of $\Graph$, and the pricing may then miss some variables.
On the other hand, if $\Omega>\Omega_d$, all edges of $\Graph'$ are also edges of $\Graph$, and therefore, the independent sets of $\Graph'$ are strictly included in the independent sets of $\Graph$. This ensures that the pricing is still able to explore all independent sets and therefore provide new variables to improve the current solution of the RMP.

As inputs, the quantum sampler gets an atom register and the pulse representing the designed Hamiltonian. Then, after running the proposed neutral atom-based quantum algorithm, different outputs are possible:

\begin{enumerate}
 \item Return only the largest independent set;
 \item Return all independent sets;
 \item Return all independent sets whose weights are greater than a given threshold;
 \item Return only the most weighted independent set;
\end{enumerate}

For applying the two last strategies, one must also provide a weight vector and a cost function as inputs to the quantum sampler. In this work, we apply the third output case mentioned above by applying the cost function~\eqref{pfo} to qualify any independent set; note that for unweighted graphs, this cost function can be applied by setting all weights to 1.

\subsection{Solving the Pricing sub-problems}
Let~$\Bar{w}$ be the vector of dual values related to the solved relaxed RMP as previously discussed. Then, we generate the graph~$\Graph' = (\vertices', \edges', \Bar{w})$ in such a way that $\vertices' =\{u \in \vertices | \Bar{w}_u > 0\}$ and~$\edges' =\{(u,v) \in \edges | \Bar{w}_u\Bar{w}_v > 0\}$. Also, for each vertex~$u \in \vertices'$, the related weight is given by~$\Bar{w}_u \in \Bar{w}$. 

%This approach is based on the quantum superposition principle. 
We run the quantum algorithm previously described on the related graph~$\Graph' = (\vertices', \edges', \Bar{w})$, where~$\Bar{w}$ is the dual vector from the current solve solution for the primal problem. While we keep only the atoms related to the vertices in~$\vertices'$, their positions might potentially be permuted as described in Algorithm~\ref{alg:permu}. Also, the pulse shape might be adjusted to the new register by calculating the new~$\Omega_{max}$, which is done by calculating the distance between each pair of qubits, as previously proposed. By applying the objective function~\eqref{pfo} to qualify each bitstring sampled by the QPU, the~$S'$ is then updated with all sampled weighted independent sets whose total weight is strictly greater than 1. Let us recall that the~$\Bar{w}$ might potentially be a vector composed of very small values (\textit{i.e.},~$\Bar{w}_u \ll 1,~\forall u \in \vertices'$, including zero and negative values), meaning that any independent set, including the maximum one, might have a total weight strictly lower than 1: in this case, no independent set is added to~$S'$. 

\begin{figure*}[t] 
 \begin{subfigure}[b]{0.329\linewidth}
 \centering
 \captionsetup{width=1\linewidth}
 \includegraphics[width=\linewidth]{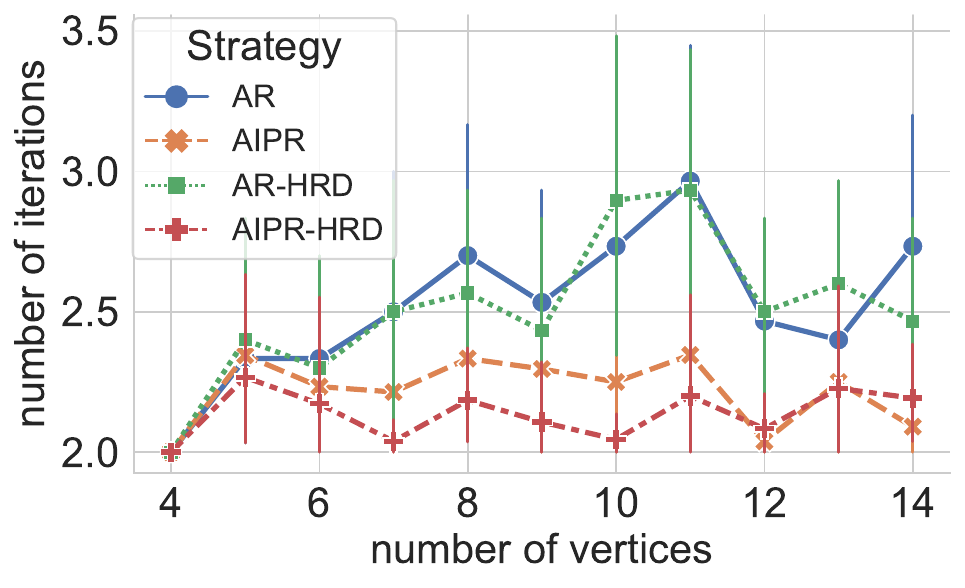} 
 \caption{}
 \label{itud02}
 \end{subfigure}
 \begin{subfigure}[b]{0.329\linewidth}
 \centering 
 \captionsetup{width=1\linewidth}
 \includegraphics[width=\linewidth]{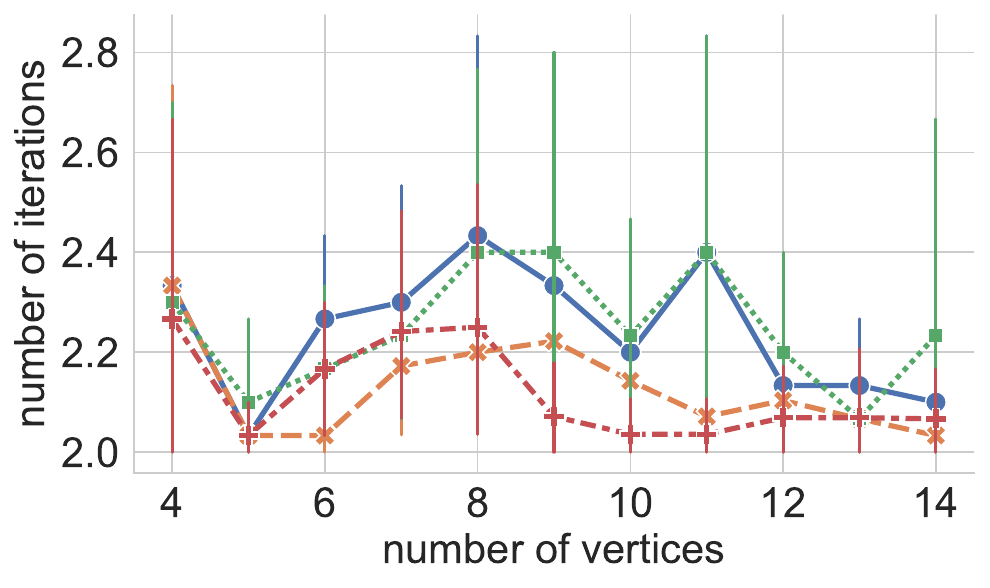} 
 \caption{}
 \label{itud05}
 \end{subfigure}
 \begin{subfigure}[b]{0.329\linewidth}
 \centering
 \captionsetup{width=1\linewidth}
 \includegraphics[width=\linewidth]{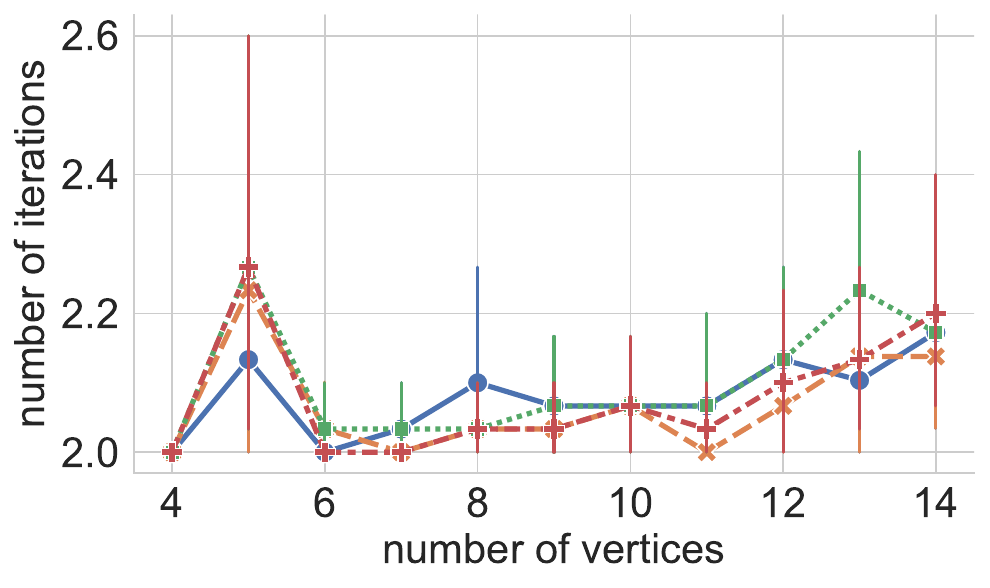} 
 \caption{}
 \label{itud08}
 \end{subfigure}
 \begin{subfigure}[b]{0.329\linewidth}
 \centering 
 \captionsetup{width=1\linewidth}
 \includegraphics[width=\linewidth]{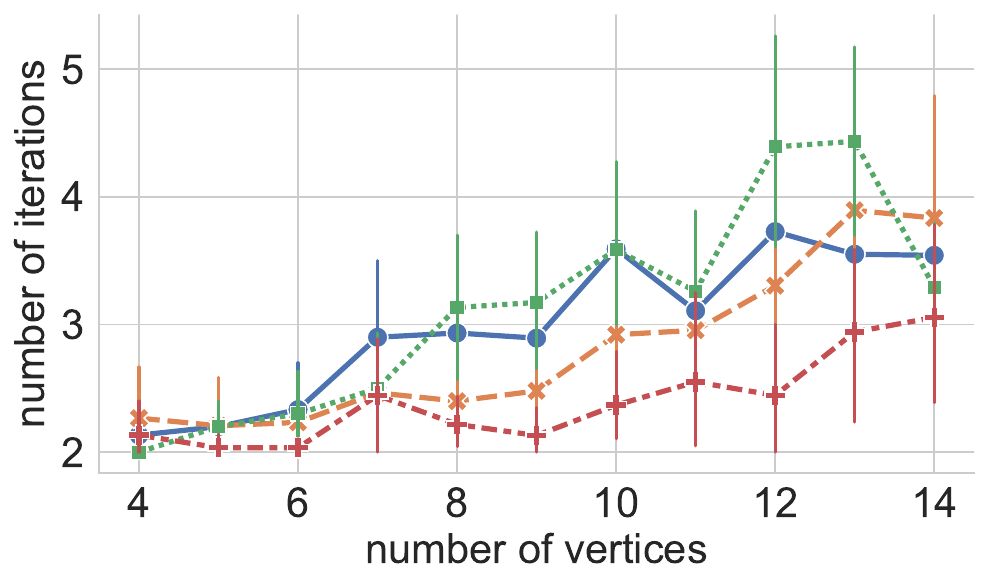} 
 \caption{}
 \label{itnud02}
 \end{subfigure}
 \begin{subfigure}[b]{0.329\linewidth}
 \centering
 \captionsetup{width=1\linewidth}
 \includegraphics[width=\linewidth]{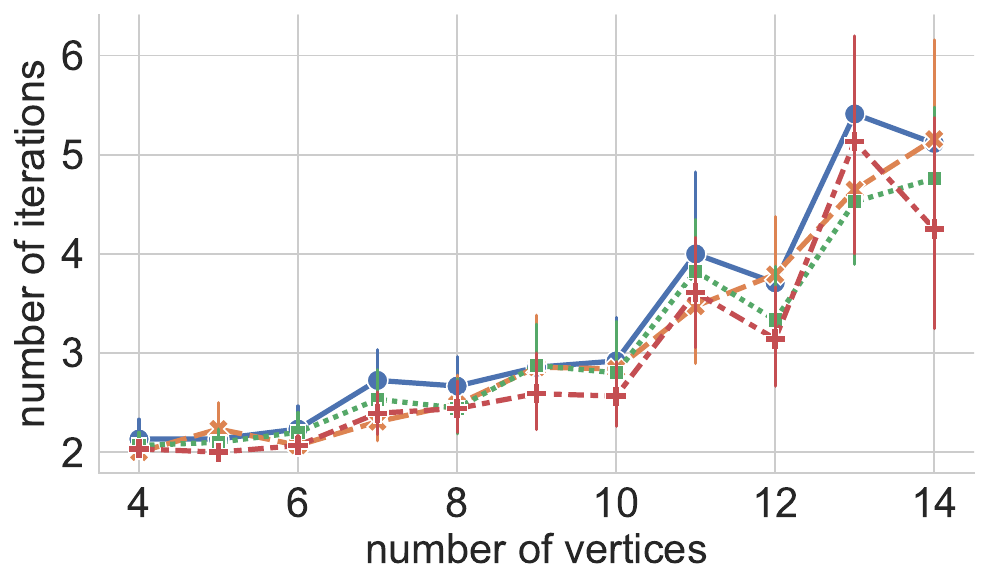} 
 \caption{}
 \label{itnud05}
 \end{subfigure}
 \begin{subfigure}[b]{0.329\linewidth}
 \centering
 \captionsetup{width=1\linewidth}
 \includegraphics[width=\linewidth]{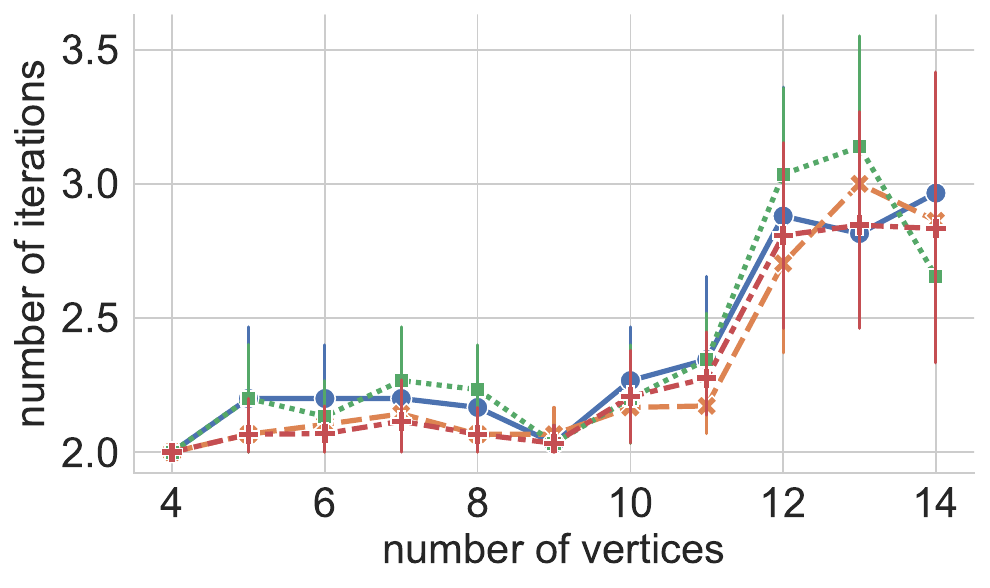} 
 \caption{}
 \label{itnud08}
 \end{subfigure} \captionsetup{justification=Justified}
 \caption{Number of iterations on the QPU emulator before reaching the stop criteria on different graph classes (UD and non-UD), orders (from 4 up to 14 vertices), and densities (20\%, 50\%, and 80\% of all possible connections) and applying different register and Hamiltonian redesign strategies: only atom removal (AR), atom index permutation and atom removal (AIPR), atom removal with Hamiltonian redesign (AR-HDR), and AIPR with Hamiltonian redesign (AIPR-HDR). Figures~\ref{itud02}-\ref{itud08} (resp. \ref{itnud02}-\ref{itnud08}) show the results on unit-disk (resp. non-UD) graphs. Also, Figures~\ref{itud02} and \ref{itnud02} (resp. \ref{itud05} and \ref{itnud05}) depict results on 20\%-density (resp. 50\%-density) graphs. Finally, Figures~\ref{itud08} and \ref{itnud08} present results on 80\%-density graphs.
} \label{itstra}
\end{figure*}

\subsection{Stopping criterion}
The algorithm stops the inner loop when no column is generated after solving the related pricing sub-problem. This means that either the current sub-set~$S'$ is already composed of all independent sets needed to find the optimal for the relaxed formulation~\eqref{rfoe}-\eqref{rce} (if the PSP is exactly solved), or the PSP solver does not find any solution under the imposed constraints (if the PSP is solved heuristically). Once the inner loop stops, the ILP version of the final RMP is solved with all generated columns (\textit{i.e.}, variables/independent sets) and the integrality constraints~\eqref{cei}.

It is important to notice that the final solution given by the related LP formulation~\eqref{rfoe}-\eqref{rce} might not be the same (or even the optimal one) for the ILP formulation~\eqref{foe}-\eqref{cei} (\textit{i.e.}, with the integrality constraints~\eqref{cei}). To ensure optimality, the proposed Hybrid Column Generation should be embedded into a Branch-and-Price framework. This approach, however, is out of the scope of this work. For more information, one may refer to~\cite{bibitem_gera1}.

\section{Numerical Simulations}\label{results}

Let us first describe the setup used in our numerical simulations. While random graph instances were generated with the Vladimir-Brandes algorithm~\cite{batagelj2005efficient}, which produces Erd\H{o}s–R\'{e}nyi graphs, UD-guaranteed graphs were produced as proposed in~\cite{penrose2003random}. It is worthwhile to mention that random graphs are unlikely to be unit-disks. Indeed, the probability of having a UD Erd\H{o}s–R\'{e}nyi graph quickly approaches zero as the number of the vertices increases and the density remains stable. For this reason, this graph class ({\sl \textit{i.e.}}, random graphs) is hereafter referred to as \textit{non-UD graphs}. For each graph class, we generated 30 instances for three different graph densities by setting the probability $p$ of connecting any pair of vertices with an edge to 20\%, 50\%, and 80\%. Also, the optimal solution of each instance was found by exactly solving the compact formulation of the Minimum Vertex Coloring problem~\cite{malaguti2010survey}.

We compare our proposed hybrid algorithm to several state-of-the-art approaches, both classical and quantum. Indeed, it is possible to solve the pricing problem~\eqref{pfo}-\eqref{pv} with a classical solver, where $\Graph'$ is given as a vertex-weighted graph. Note, however, that only the optimal weighted independent set is generated on each iteration. As previously discussed, due deterministic nature of this approach, hereafter referred to as \textit{Classical CG}, the final solution does not change same if the input (i.e., the graph and vertex weights) remains the same. We also compared our proposed approach to the \textit{Classical Greedy} algorithm described in Appendix \ref{greedyyy}, where a maximum independent set was provided by a classical solver on each iteration. We use GLPK~\cite{glpk} as the linear solver to solve the (reduced) master problem and classical pricing sub-problems.

In order to compare our neutral atom-based quantum sampler to other stochastic approaches, we implemented a greedy generator, hereafter referred to as \textit{Greedy CG}, that can randomly generate multiple weighted independent sets. For more details, see Appendix~\ref{greedypricing}. 
We also compared our approach to a simulated annealing (SA)-based solver,  hereafter referred to as \textit{SA CG}, where we minimize the related maximum weighted independent set QUBO matrix described in \eqref{eq:misqubo} with the classical D-Wave QUBO sampler \cite{dwavesampler}; the weights are given by the dual values of the solved RMP on each iteration, while $\alpha$ is set to the sum of absolute values of weights. For all stochastic sub-routines, the maximum number of tries was set to 1000. Moreover, several pricing iterations can be done before getting an improvement on the RPM solution. This is due to the inherent symmetry of the solution space related to the instance of the problem, driving the algorithm to generate independent sets with the same cost. In all CG-based approaches, the maximum number of pricing iterations allowed without any improvements in the RMP solution was set to 3. Finally, we also compared our hybrid approach against a quantum version of the greedy algorithm described in Appendix~\ref{greedyyy}, where only the largest independent set sampled by the proposed quantum sampler described in section~\ref{quantumsampler} is returned on each iteration. This approach is hereafter referred to as \textit{Quantum Greedy} and is based on the work proposed by Vitali \textit{et al}~\cite{21:sc:quantum}.

\begin{table}[b]\footnotesize
 \centering
 \begin{tabular}{|c|c|c|}\hline
 \multirow{3}[1]{*}{~~SPAM~~} & bad preparation $\eta$ & 0.005 \\
\cline{2-3} & false positive  $\epsilon$ & 0.03 \\
\cline{2-3} & false negative  $\epsilon'$ & 0.08 \\\hline
 \multicolumn{2}{|c|}{~~Temperature~~} & $30 \mu K$ \\\hline
 \multicolumn{2}{|c|}{Laser waist} & ~~$148 \mu m$~~ \\\hline
 \end{tabular}
 \caption{Noise parameters used in \textit{noisy} emulations.
 %SPAM errors are parametrized by three probabilities $\eta$, $\epsilon$ and $\epsilon'$. The optical pumping process that prepares all the atoms in the $\ket{0}$ state might fail with probability $\eta$, resulting in some atoms not participating in the quantum evolution at all. 
 }
 \label{tab:noise}
\end{table}

%Quantum environment 
%All numerical simulations of the quantum device were designed and run on \textit{Pulser}~\cite{silverio2022pulser}, an open-source python library for programming neutral-atom devices at the pulse level. % with high fidelity. In Pulser, a pulse can be built by specifying two time-dependent waveforms: one for the Rabi frequency of the laser and one for the detuning. Each waveform was obtained by interpolating between five free points equally spaced along the pulse duration. Fixing the initial and final value of the Rabi frequency to zero gives a total of nine free parameters: three for $\Omega$, five for $\Delta$, and one for the pulse duration. %The resulting waveforms are of the type shown in Fig.~\ref{fig:easy_pulse}, where the round markers correspond to the points between which the curve is interpolated. The parameters were bound by realistic hardware specifications. The

\begin{figure*}[t] 
 \begin{subfigure}[b]{0.329\linewidth}
 \centering
 \captionsetup{width=1\linewidth}
 \includegraphics[width=\linewidth]{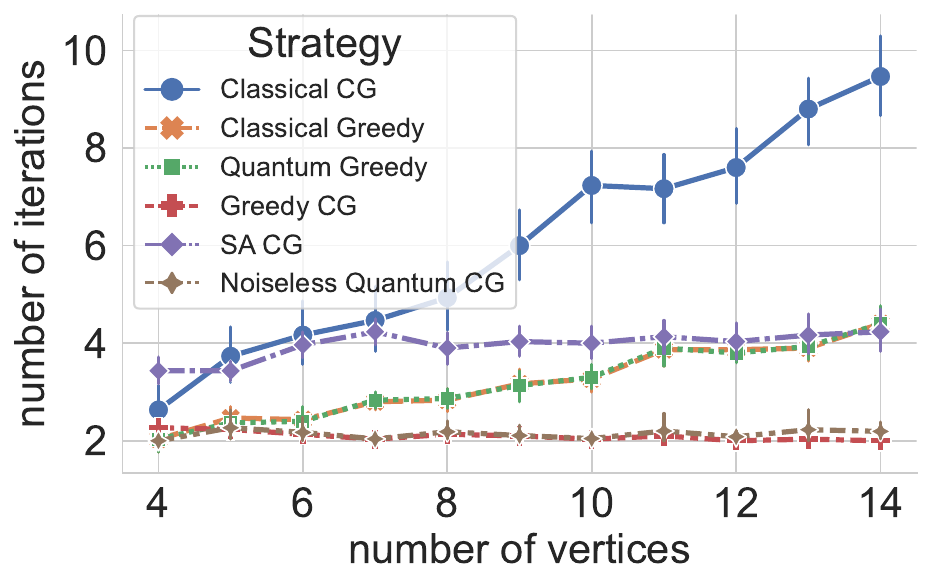} 
 \caption{}
 \label{apitud02}
 \end{subfigure}
 \begin{subfigure}[b]{0.329\linewidth}
 \centering 
 \captionsetup{width=1\linewidth}
 \includegraphics[width=\linewidth]{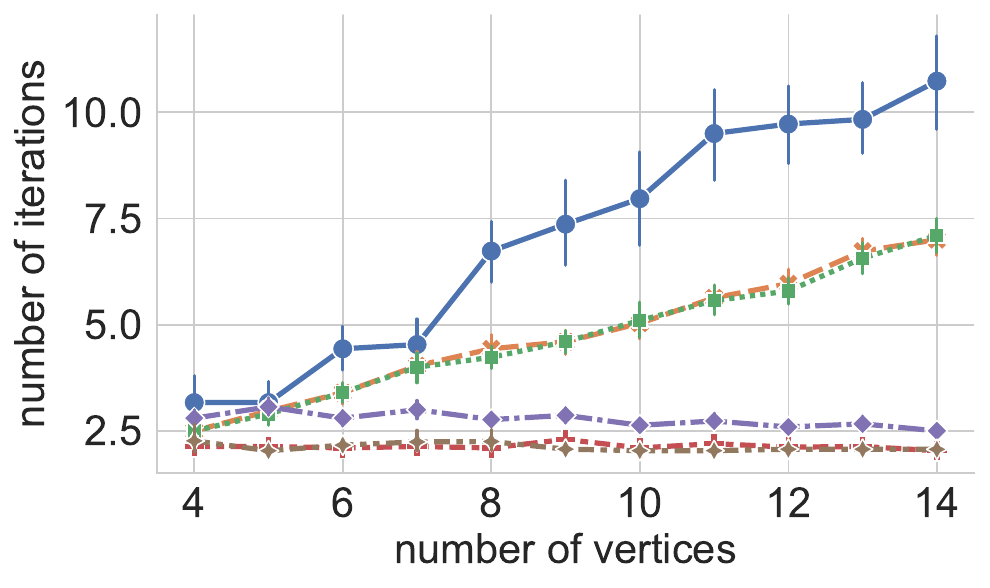} 
 \caption{}
 \label{apitud05}
 \end{subfigure}
 \begin{subfigure}[b]{0.329\linewidth}
 \centering
 \captionsetup{width=1\linewidth}
 \includegraphics[width=\linewidth]{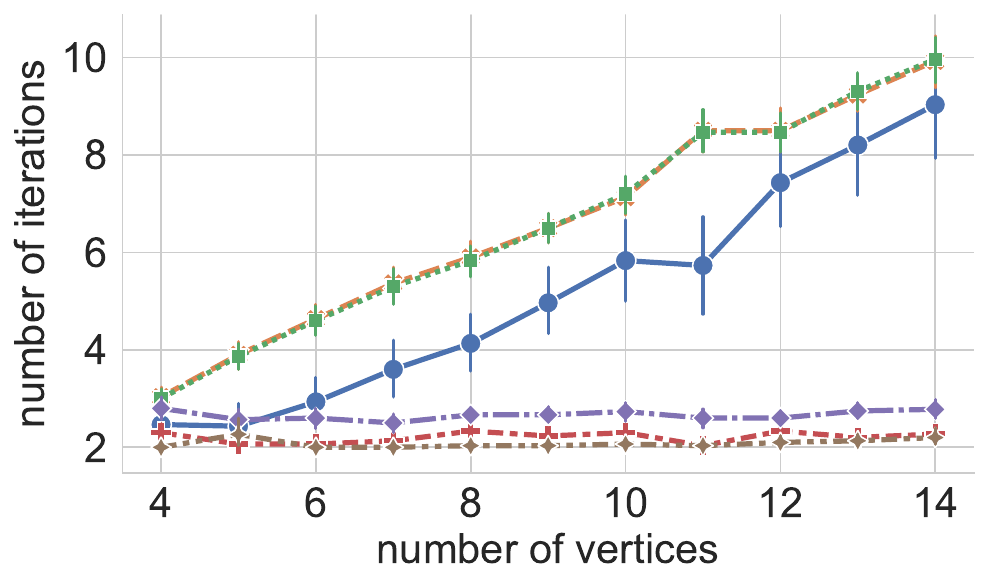} 
 \caption{}
 \label{apitud08}
 \end{subfigure}
 \begin{subfigure}[b]{0.329\linewidth}
 \centering 
 \captionsetup{width=1\linewidth}
 \includegraphics[width=\linewidth]{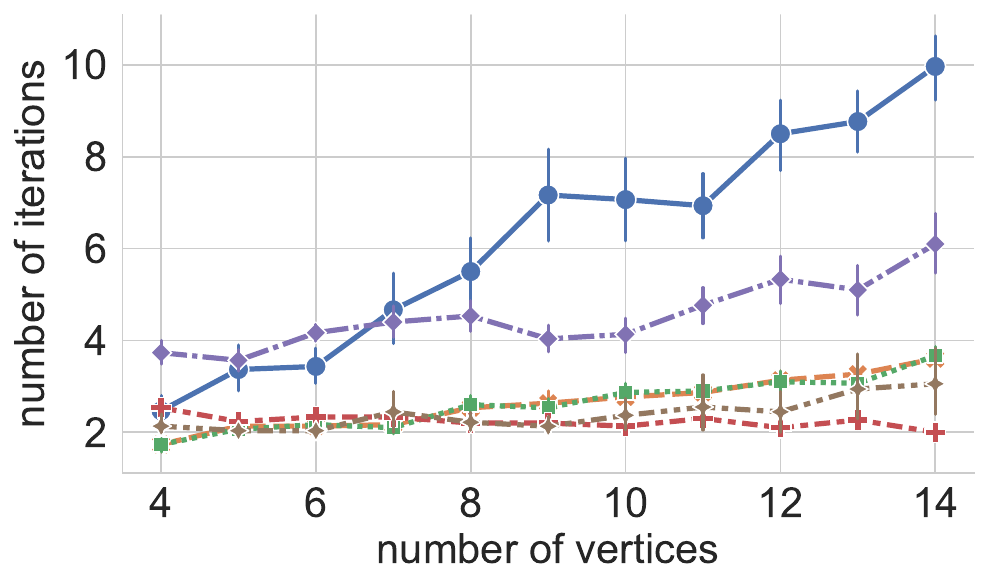} 
 \caption{}
 \label{apitnud02}
 \end{subfigure}
 \begin{subfigure}[b]{0.329\linewidth}
 \centering
 \captionsetup{width=1\linewidth}
 \includegraphics[width=\linewidth]{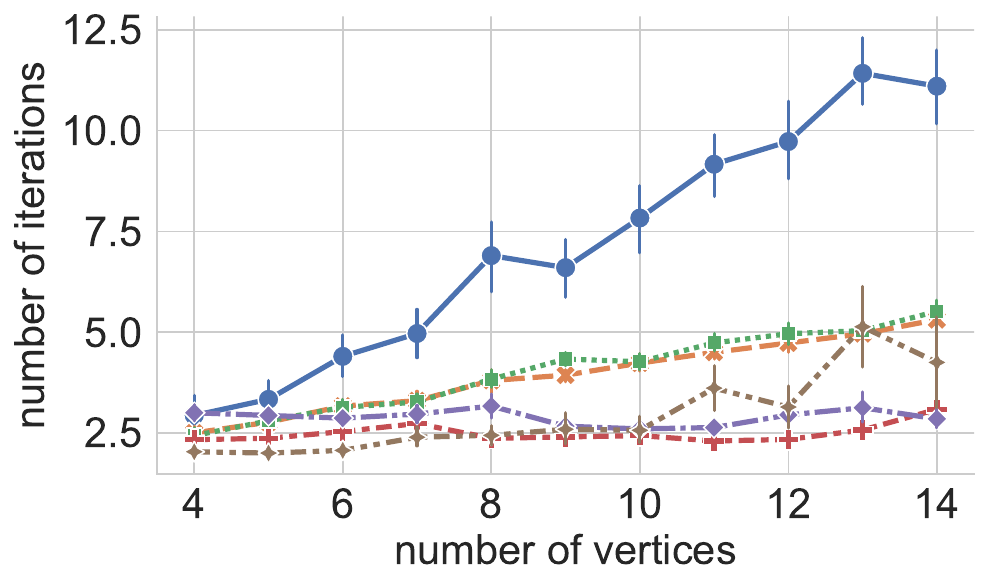} 
 \caption{}
 \label{apitnud05}
 \end{subfigure}
 \begin{subfigure}[b]{0.329\linewidth}
 \centering
 \captionsetup{width=1\linewidth}
 \includegraphics[width=\linewidth]{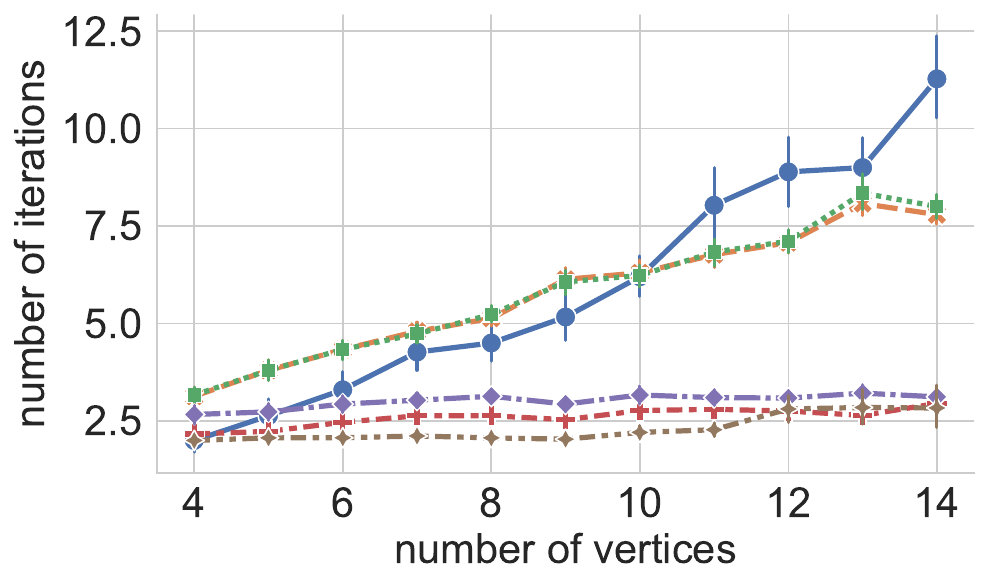} 
 \caption{}
 \label{apitnud08}
 \end{subfigure} \captionsetup{justification=Justified}
 \caption{Number of iterations before reaching the stop criteria on different graph classes (UD and non-UD), orders (from 4 up to 14 vertices), and densities (20\%, 50\%, and 80\% of all possible connections) by applying different approaches: Classical Column Generation (CG), Greedy CG, SA CG, Quantum CG, Classical Greedy, and Quantum Greedy. Figures~\ref{apitud02}-\ref{apitud08} (resp.~\ref{apitnud02}-\ref{apitnud08}) show the results on unit-disk (resp. non-UD) graphs. Also, Figures~\ref{apitud02} and \ref{apitnud02} (resp.~\ref{apitud05} and \ref{apitnud05}) depict results on 20\%-density (resp.~50\%-density) graphs. Finally, Figures~\ref{apitud08} and \ref{apitnud08} present results on 80\%-density graphs.}
 \label{itappro}
\end{figure*}

The register related to each pricing sub-problem was created by applying the embedding strategy presented in Section~\ref{spring_section}. To this end, each atom's position was found with the \textit{spring layout} function from \textit{Networkx} package~\cite{hagberg2008exploring}: we multiplied each position vector by 40 in order to respect the distance constraints imposed by the device. Moreover, we applied the Hamiltonian design strategy described in Section~\ref{quantumsampler}. 

The evolution of the quantum system under the predicted pulse for a given register was then simulated using Pulser's simulation module~\cite{silverio2022pulser}. Both noiseless and noisy simulations were performed. Noiseless simulations involve solving the time-dependent Schr\"{o}dinger equation. The output of a noiseless simulation is a vector in the Hilbert space that can be sampled a finite number of times in order to mimic a real experimental setup with a limited measurement budget.
In order to assess the robustness of our approach against noise, noisy simulations taking into account the current levels of noise in these systems~\cite{cacib, qeqq} were also performed.
These calculations are numerically expensive and we restrict our study to State Preparation And Measurement (SPAM) errors.
These are expected to be the main source of noise and can be obtained by post-processing of noiseless results (for more details, see~\cite{qscore}).
We set here the noise parameters to realistic values based on current hardware specifications. They are summarized in Table~\ref{tab:noise}.  The register and Hamiltonian design were done as previously presented in sections \ref{spring_section} and \ref{quantumsampler}, respectively.

%, on the other hand, can be rather cumbersome depending on the type of noise to be included.
%The least expensive noise source is related to measurement errors $\epsilon$ and $\epsilon'$, and can  any kind of state sampling.
%Including preparation errors, laser defects, and temperature effects would require, in principle, performing a new simulation for each sample that has to be collected.
%To find a compromise between computational resources and simulation accuracy, we decided to perform only five independent noisy simulations and to collect 10 samples from each. The last noise type available in Pulser, the dephasing channel accounting for spontaneous emission, would force the adoption of the density matrix formalism and the solution of the Lindblad master equation~\cite{manzano2020short}, introducing a rather severe computational overhead. For this reason, and because it is expected to have a smaller effect for short pulses, noisy simulations included in this work did not take dephasing into account. One set of noisy emulations was then performed, whose noise parameters are summarized in Table~\ref{tab:noise} and is based on current hardware specifications.

\subsection{Register and pulse redesign strategies}
We first analyze the impact of four different registers and pulse (\textit{i.e.}, Hamiltonian) redesign strategies on the performance of our proposed hybrid classical-quantum column generation approach: while the Atom Removal (AR) strategy only removes the atoms whose related vertex's weight is equal to or less than zero, Atom Index Permutation and Atom Removal (AIPR) strategy also apply the proposed vertex-atom remapping algorithm (see Algorithm~\ref{alg:permu}). In addition to these two strategies, we tested the Atom Removal with Hamiltonian Redesign (AR-HDR), in which the new maximal Rabi frequency is recalculated after applying the AR strategy. Finally, in Atom Index Permutation, Atom Removal and Hamiltonian Redesign (AIPR-HDR) strategy, the new maximal Rabi frequency can also be recalculated after applying the AIPR strategy.

Fig.~\ref{itstra} shows the average number of pricing iterations (and the standard deviation with 95\% confidence interval) before reaching the stop criteria on different graph classes (UD and non-UD), order (from 4 up to 14 vertices), and densities (20\%, 50\%, and 80\% of all possible connections). As observed, only a few calls to the quantum sampler (\textit{i.e.}, pricing iterations) were needed to reach the best solution of the relaxed version\footnote{Let us recall that the dual variables can be only generated from linear programs, meaning that the integrality constraints~\eqref{cei} are replaced by constraints~\eqref{rce}.} of the master problem related to each graph instance. Indeed, the average number of iterations on non-UD (resp.~UD) graphs was always less than 6 (resp.~3); as seen in Fig.~\ref{itud08} (resp.~Fig.~\ref{itnud08}), only 2 (resp.~3) sampling processes were done in average to solve UD (resp.~non-UD) graphs with 4 (resp.~13) vertices and 80\% density when AIRP (resp.~AR-HDR) strategy was applied. Even though non-UD graphs seem to be more difficult to be solved, especially those with~50\% of density (see Fig.~\ref{itnud05}), applying different register and pulse redesign strategies could speed up the solving process. While we do not observe any significant impact from redesigning the pulse after only removing useless atoms from the register (see blue and green lines respectively related to AR and AR-HDR strategies), recalculating the maximum value of the Rabi frequency after permuting atoms' indices (\textit{i.e.}, applying AIPR-HDR approach) could decrease the number of sampling processes by~44\% (see Fig.~\ref{itnud02}). Indeed, as seen in Figures~\ref{itud02}~and~\ref{itnud02}, this approach had the best overall performance, having a stronger impact on sparse~graphs.

\begin{figure*}[t] 
 \begin{subfigure}[b]{0.329\linewidth}
 \centering
 \captionsetup{width=1\linewidth}
 \includegraphics[width=\linewidth]{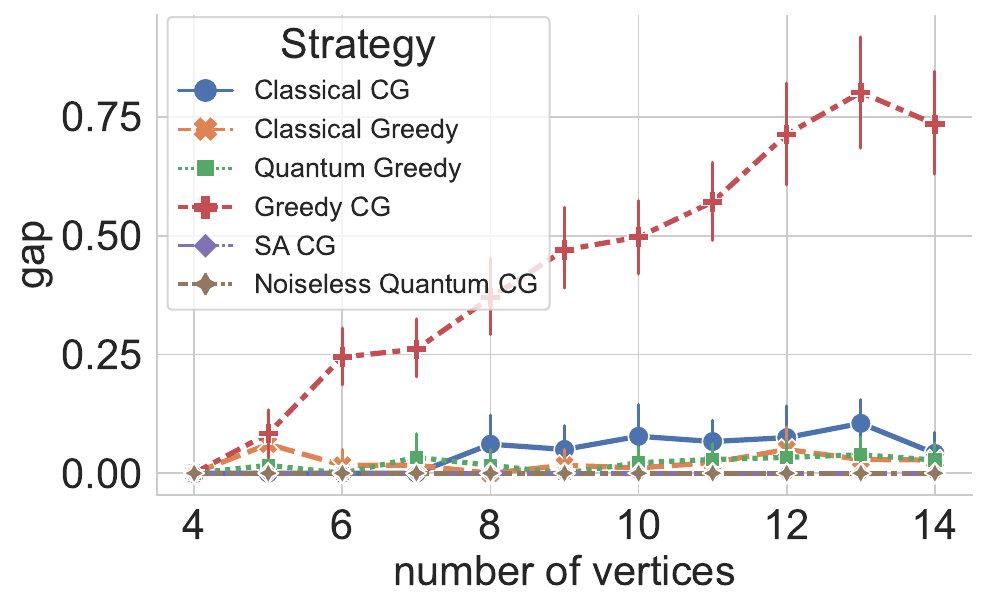} 
 \caption{}
 \label{gapud02}
 \end{subfigure}
 \begin{subfigure}[b]{0.329\linewidth}
 \centering 
 \captionsetup{width=1\linewidth}
 \includegraphics[width=\linewidth]{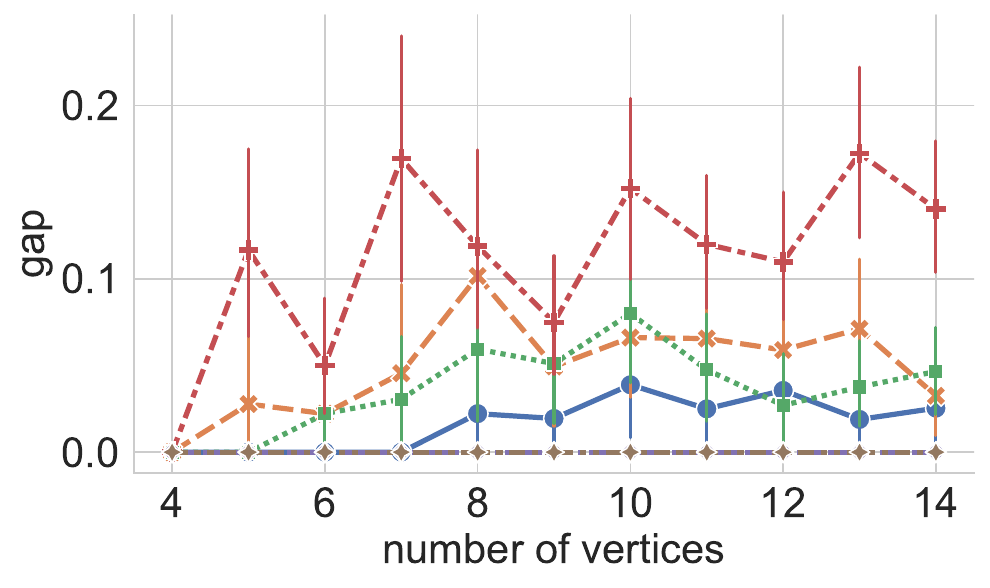} 
 \caption{}
 \label{gapud05}
 \end{subfigure}
 \begin{subfigure}[b]{0.329\linewidth}
 \centering
 \captionsetup{width=1\linewidth}
 \includegraphics[width=\linewidth]{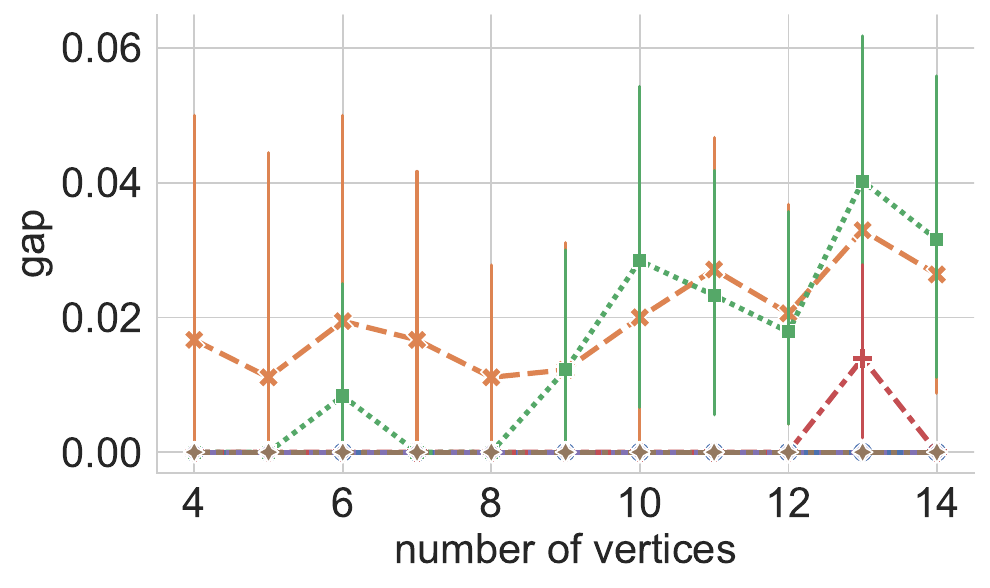} 
 \caption{}
 \label{gapud08}
 \end{subfigure}
 \begin{subfigure}[b]{0.329\linewidth}
 \centering 
 \captionsetup{width=1\linewidth}
 \includegraphics[width=\linewidth]{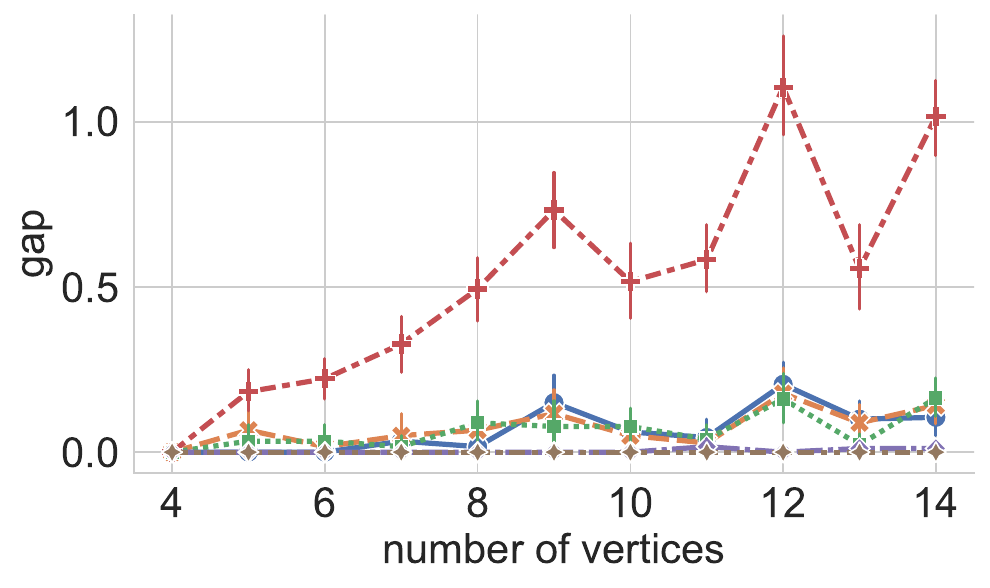} 
 \caption{}
 \label{gapnud02}
 \end{subfigure}
 \begin{subfigure}[b]{0.329\linewidth}
 \centering
 \captionsetup{width=1\linewidth}
 \includegraphics[width=\linewidth]{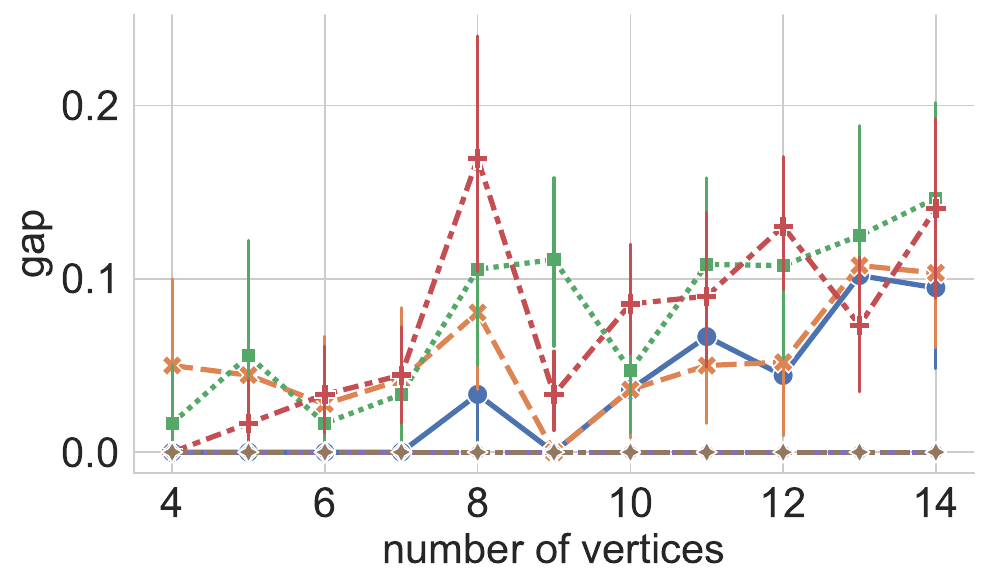} 
 \caption{}
 \label{gapnud05}
 \end{subfigure}
 \begin{subfigure}[b]{0.329\linewidth}
 \centering
 \captionsetup{width=1\linewidth}
 \includegraphics[width=\linewidth]{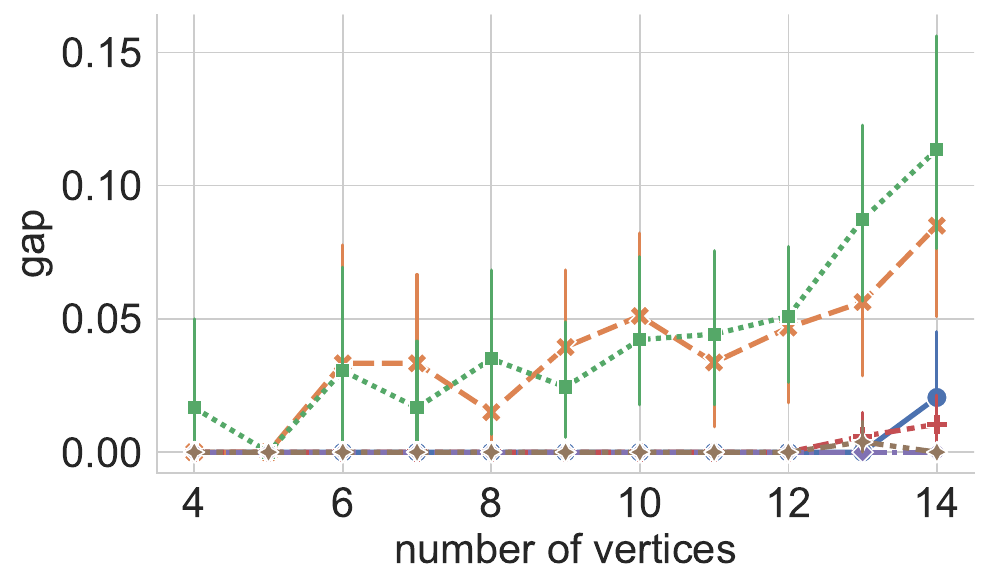} 
 \caption{}
 \label{gapnud08}
 \end{subfigure} \captionsetup{justification=Justified}
 \caption{Gab between the best solution found and the optimal one on different graph classes (UD and non-UD), orders (from 4 up to 14 vertices), and densities (20\%, 50\%, and 80\% of all possible connections) by applying different approaches: Classical Column Generation (CG), Greedy CG, SA CG, Noiseless Quantum CG, Classical Greedy, and Quantum Greedy. Figures~\ref{gapud02}-\ref{gapud08} (resp.~\ref{gapnud02}-\ref{gapnud08}) show the results on unit-disk (resp. non-UD) graphs. Also, Figures~\ref{gapud02} and \ref{gapnud02} (resp.~\ref{gapud05} and \ref{gapnud05}) depict results on 20\%-density (resp.~50\%-density) graphs. Finally, Figures~\ref{gapud08} and \ref{gapnud08} present results on 80\%-density graphs. }
 \label{gapappro}
\end{figure*}
\subsection{Quantum and classical approaches}
We now compare the number of iterations needed to be run on different graph classes, orders, and densities by applying different approaches: Classical Column Generation (CG), Greedy CG, SA CG, Noiseless Quantum CG, Classical Greedy, and Quantum Greedy. We applied the AIPR-HDR strategy for redesigning each pricing sub-problem within the Quantum CG framework. While this indicator refers to how many times the PSP was solved within both classical and quantum column generation frameworks for coloring a given graph, it indicates how many independent sets were generated during the while-loop on Algorithm~\ref{alg:greed} by using both classical and quantum methods as previously discussed. Fig.~\ref{itappro} shows the average number of iterations and the standard deviation (with 95\% confidence interval) on 30 graphs randomly generated as presented above.

First, we observe that the Quantum Greedy approach has the same overall performance as its classical counterpart, showing that our quantum sampler can solve the Maximum Independent Set problem efficiently.  Also, both strategies have the same linear behavior related to the size of the graph, \textit{i.e.}, the number of edges it contains, being most impacted by dense graphs (see Figures~\ref{apitud08} and~\ref{apitnud08}). This behavior is expected since the size of each independent set gets smaller as the set of edges gets larger. Hence, more iterations have, in general, to be done to cover all vertices of a dense graph.  Also, while outperforming the Classical CG approach on almost every graph class (in terms of the number of iterations), both Classical and Quantum Greedy algorithms had their performance slightly decreased on UD graphs. Finally, taking advantage of the related superposition aspect, the proposed Quantum CG outperformed all other approaches on all graph classes. For instance, while the Quantum CG algorithm needed less than 4 (resp.~6) sampling iterations for all sparse and dense (resp.~0.5-density) non-UD graphs, Quantum and Classical Greedy approaches (resp.~Classical CG algorithm) needed up to 10 (resp.~12) pricing interactions to solve the same graph class.

Fig.~\ref{gapappro} shows the average gap\footnote{Throughout the paper, the gap is calculated as follows: for a given problem instance and the value $x_o$ of the optimal solution (i.e., the minimum number of colors needed to color the graph, also known as the chromatic number of the graph) and the value $x_m$ of the solution (i.e., the number of colors used to color the graph) after running a given method the gap is given by $\dfrac{x_o - x_m}{x_o}$. In other words, the gap represents the distance between a given solution found by any method from the optimal one.} (and the standard deviation with 95\% of confidence interval) between the optimal solution and the best one found by applying the presented approaches. First, we observe that our proposed Quantum CG approach has the best overall performance. Indeed, it could find the optimal solution in almost all instances; as seen in Fig.~\ref{gapnud08}, our approach could not find the best solution only for some 13-vertex non-UD graphs. Also, unlike all other approaches, the Quantum CG is not impacted by the graph class; while the Classical CG could better perform on dense graphs (see Figures~\ref{gapud08} and~\ref{gapnud08}), both Classical and Quantum Greedy approaches are more stable on UD graphs. Also, as depicted in Fig.~\ref{gapnud02} (resp. Fig.~\ref{gapud02}), the proposed Quantum CG algorithm could reduce the average gap on 12-vertex non-UD (resp. 13-vertex UD) graphs from roughly 19\% (resp. 11\%) to 0\% when compared to the Quantum Greedy (resp. Classical CG) approach.

Our proposed quantum pricing-based approach also outperformed both stochastic classical approaches in most of the instances, especially those related to UD and sparse graphs. Even though the quality of the solutions remains the same (see Fig.~\ref{gapappro}), the Noiseless Quantum CG could reduce by 50\% the number of iterations on sparse graphs when compared to SA-based pricing, as observed in Figures~\ref{apitud02} and~\ref{apitnud02}). Even though the Greedy CG has fewer pricing iterations on some graph classes, as in bigger non-UD graphs with 20\% and 50\% of density (see Figures~\ref{apitnud02} and \ref{apitnud05}), the average gap could be reduced by 80\% when our proposed Quantum CG was applied on the same graph classes, as we observe in Figures~\ref{gapud02} and~\ref{gapnud02}. This indicates that random sampling to find independent sets cannot solve pricing sub-problems effectively.  

\subsection{Noisy and noiseless simulations}

\begin{figure}[b] 
 \begin{subfigure}[b]{0.97\linewidth}
 \centering
 \captionsetup{width=1\linewidth}
 \includegraphics[width=\linewidth]{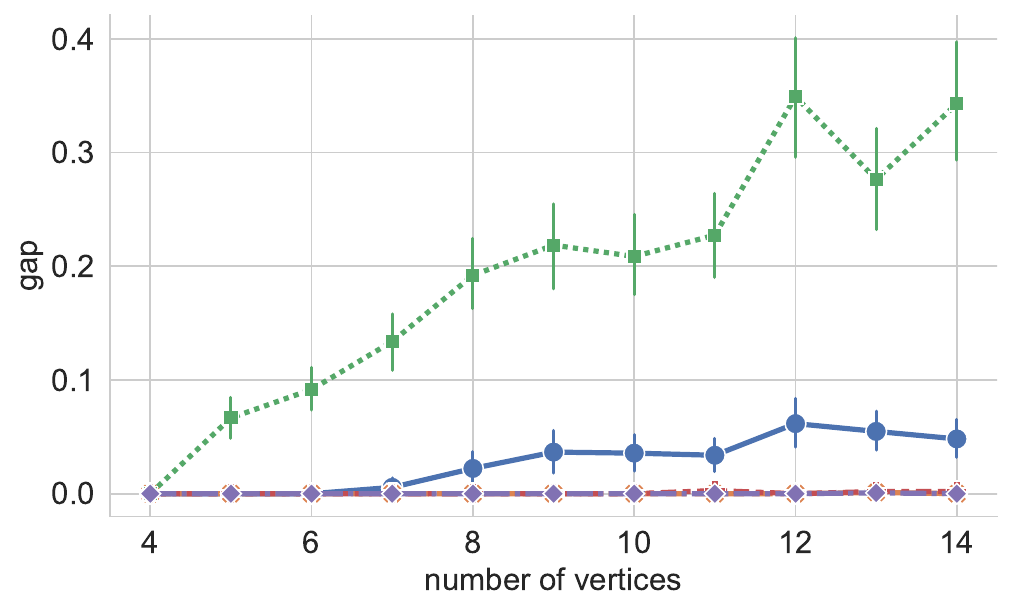} 
 \caption{}
 \label{gapnoisy}
 \end{subfigure}
 \begin{subfigure}[b]{0.97\linewidth}
 \centering 
 \captionsetup{width=1\linewidth}
 \includegraphics[width=\linewidth]{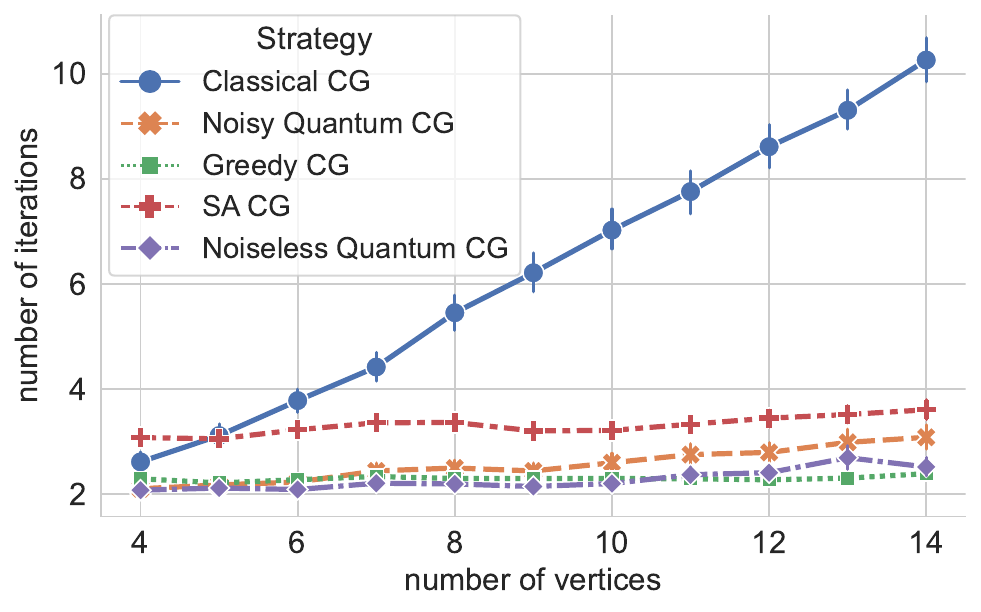} 
  \caption{}
 \label{itnoisy}
 \end{subfigure}\captionsetup{justification=Justified}
 \caption{Number of iterations before reaching the stop criteria and the gap between the final solution and the optimal one on different graph orders (from 4 up to 14 vertices) applying different approaches: Classical Column Generation~(CG), Greedy CG, SA CG, Noiseless Quantum CG, and Noisy Quantum CG. While Fig. \ref{gapnoisy} presents the gap between the final solution and the optimal one, Fig. \ref{itnoisy} shows the number of iterations before reaching the stop criteria.}
 \label{itgapnose}
\end{figure}

We now present the results from our numerical simulations wherein emulated noise was added as previously discussed. Fig.~\ref{itgapnose} depicts the number of iterations before reaching the stop criteria and the gap between the final solution and the optimal one on different graph orders (from 4 up to 14 vertices) by applying different approaches: Classical CG, Greedy CG, SA CG,  Noiseless Quantum CG, and Noisy Quantum CG. The results shown are related to applying a given approach on all graph instances of the same order, regardless of their density and whether they are unit disks. First, as seen in Fig.~\ref{gapnoisy}, no impact was observed when noise is added to the quantum independent set sampler. Indeed, the overall final gap was similar to the noiseless model, also outperforming the classical CG. Finally, as observed in Fig.~\ref{itnoisy} the number of iterations needed to find the optimal solution of the relaxed RMP was increased by only 6\% on average when the noisy model is compared to the noiseless one, even on big graphs.
  
\section*{Conclusion}
In this study, we demonstrated that it is possible to incorporate quantum elements into classical state-of-the-art algorithms in order to improve their performances. Compared to the classical column generation, our neutral atom-based quantum pricing could reduce by up to 83\% the number of iterations needed to solve the Minimum Vertex Coloring problem. Also,  our proposed hybrid approach could reduce the average gap from 19\% to 0\%  when compared to both classical and greedy approaches. Moreover, unlike the deterministic approaches, our quantum pricing-based column generation is robust to all graph classes, including non unit-disk graphs of all tested orders and sizes. This indicates that the proposed hybrid algorithm can efficiently solve other combinatorial problems (e.g., Minimum Edge Coloring and Minimum Clustering problems) after some trivial translation-based pre-processing. The proposed framework also outperformed other stochastic algorithms, especially on sparse graphs. Indeed, our quantum pricing-based column generation could reduce by up to 50\% the number of pricing iterations and by up to 80\% the gap of the final solution on graphs when compared to the simulated annealing-based and greedy-based pricing approaches, respectively. Finally, we observed that our proposed hybrid framework is robust to noise. Indeed, the quality of the final solutions was not impacted when compared to the noiseless model, which is a great indication of how noise-resilient the analog mode of operation can be. Our proposed approach can readily be implemented on neutral-atom quantum computing hardware.

Let us recall that our proposed algorithm remains a heuristic approach to solving combinatorial problems. Even though it can provide high-quality solutions to a plurality of instance classes, embedding this method into a Branch-and-Pricing framework is necessary to guarantee optimality to any instance of the problem under consideration. Since providing the initial sub-set of variables for the reduced master problem is an important step in any column generation-based algorithm, the proposed quantum sampler can be used as a warm starter to generate such a subset of elements (e.g., independent sets). In the case of solving the Minimal Vertex Coloring problem, by setting all vertex weights to 1, this approach might be useful in scenarios where QPU resources are limited. Also, this strategy might potentially speed up both classical and quantum column generation approaches. Moreover, an optimal control-based strategy to design pulses to each input instance might potentially be applied to solve non-trivial pricing sub-problems. Similarly, different register embedding approaches can be developed to take into consideration different information from the input data. 

\section*{Acknowledgments}
We thank Julien Bernos and Vincent Elfving for insightful discussions. During the completion of this manuscript, we became aware of a related work\,\cite{vesh}.

\appendix

\section{A Greedy algorithm for the Minimum Vertex Coloring problem}\label{greedyyy}
We present here a greedy heuristic based on the Minimum Vertex Coloring framework introduced by~\cite{21:sc:quantum}. The main idea relies on interactively solving the Maximum Independent Set problem with only a subset of the vertices of a given graph. Algorithm~\ref{alg:greed} summarizes the proposed approach. 

\begin{algorithm}[!h]
\caption{Greedy Algorithm}\label{alg:greed}
\begin{algorithmic}[1]
 \Require A Graph $\Graph = (\vertices, \edges)$ and a set $C$ of available colors.
 \Ensure Color-vertex assignment.
\State $\Graph' \gets G$
\State $c\gets 1$ 
\While{$\Graph'$ has vertices}
\State Find a (maximum) independent set $IS$ in $\Graph'$
\State Assign color $c\in C$ to all vertices from $IS$ in $\Graph$
\State Remove all incident edges of each vertex in $IS$ from $\Graph'$.
\State Remove the set $IS$ of vertices from $\Graph'$. 
\State $c = c + 1$ 
\EndWhile
\end{algorithmic}
\end{algorithm}

As input, the Algorithm~\ref{alg:greed} receives a graph $\Graph = (\vertices,\edges)$ and a set $C$ of available colors. Note that, to always have a feasible color assignment, $|\vertices| \leq |C|$ must hold. First, a copy of $\Graph$ is made with an auxiliary graph $\Graph'\equiv~G$. A variable $k$ is also created: it keeps the index of the first available color. Then, steps~3-7 are done until no vertex remains in~$\Graph'$. On each iteration, a feasible independent set in $\Graph'$, potentially a maximal one, is generated. For instance, a classical solver can be used to solve the formulation~\eqref{eq:misqubo} or one may use the proposed quantum sampler described in section~\ref{quantumsampler} by setting the algorithm to output only the largest sampled independent set.
Then, all vertices of G with the same indices as those in the found independent set $IS$ are colored with the first available color from $C$, whose index is given by the variable $c$. Next, all vertices of $IS$ (and the related incident edges) are removed from $\Graph'$, and then the reference of the first available color is updated (see step~8). 

It is worth mentioning that, even though the number of qubits needed to run this algorithm on a QPU is reduced when compared to other approaches\footnote{While some QUBO formulations for the Vertex Coloring Problem present $|V|^2$ variables, and hence at least $|V|^2$ qubits are needed to encode their counterpart \cite{ardelean2022graph, glover2022quantum} (e.g., when no auxiliary qubit is used), the greedy algorithm presented in~\cite{21:sc:quantum} uses at most $|V|$ variables/qubits.}, its performance can be strongly impacted by the order in which the independent sets are generated. Also, in the worst case, $|\vertices|$ iterations can be done in the quantum device (take a complete graph as an example) before the final solution is found. More details about its performance are presented in Section~\ref{results}.

\section{Force-directed algorithm}\label{FRalgo}

As pointed in~\cite{pasqore}, algorithms based on force-directed principles are normally used to embed graphs into planes in such a way that two connected (resp.~disjoint) vertices are placed close to (resp.~far from) each other, with a minimum (resp.~maximum) distance between them (resp.~from the plane's center). In this context, in order to reflect inherent symmetries, Fruchterman and Reingold~\cite{fruchterman1991graph} also propose an efficient algorithm to place the vertices evenly in the plane, making the edges' lengths uniform. For this purpose, each edge from the graph is treated as a spring that holds its endpoint vertices close to each other while a competing repulsive force is applied to push all vertices away from one another, even though they are not connected by an edge in the original graph. After enough iterations, the final system will reach equilibrium, minimizing then the difference between all attractive and repulsive forces.

The repulsive and attractive forces $f_r$ and $f_a$ between two vertices are respectively given by equations~\eqref{repul} and~\eqref{attr}. While $k = \sqrt{A/|\vertices|}$ is set to be related to the area $A$ of the Euclidean plane, $d_{uv}$ holds the distance between the vertices $u,v \in \vertices$. Finally, the total energy $f_t$ of the system is given by adding the forces between all pairs of vertices, as shown in~\eqref{total}. Therefore, $f_t$ goes to zero as the system approaches its equilibrium. This register embedding is hereafter referred to as \textit{spring layout}. 

\begin{align}
 & f_r(u,v) = -k^2/r_{uv} \label{repul}\\
 & f_a(u,v) = d_{uv}^2/k \label{attr}\\
 & f_t = \sum_{u,v \in \edges}f_a(u,v) + \sum_{u \in \vertices}\sum_{v \in \vertices: u\neq v}f_r(u,v) \label{total}
\end{align}

Fig.~\ref{gtbe} depicts a possible embedding by directly applying the algorithm on a graph with 5 vertices and 7 edges. For a deep description of the algorithm, one may refer to~\cite{pasqore,fruchterman1991graph}. If such a realization exists, the graph under consideration is naturally embedded as a UD graph if enough iterations are allowed (\textit{i.e.}, by iterating until the system reaches the equilibrium).  It is also worthwhile mentioning that, as pointed out by~\cite{pasqore}, the proposed embedding strategy is only feasible on neutral atom-based QPUs once they respect the device's technical constraints, such as maximum distance from the register's center and minimum distance between atoms. If any technical constraint is violated, one might re-scaling every position vector by a factor $\alpha > 0$.

\section{Classical greedy pricing algorithm}\label{greedypricing}

Algorithm \ref{algo:greedypricing} summarizes the proposed random weighted independent set generator. As input, the algorithm receives a graph $\Graph = (\vertices, \edges, W)$, where $W$ is the vertex weighting vector. First, an auxiliary graph $\Graph'$ is created to receive a copy of the input graph $\Graph$ (step 3). Then, a vertex from graph $\Graph'$ is randomly selected and added to the independent set $S$ (see steps 6-7). Next, the selected vertex and its neighbors are removed from $\Graph'$ in step 8. Steps 6-8 are repeated until no vertex remains. The overall cost of the final solution is then calculated considering the vertex weighting vector $W$ (see step 11). Due to the inherently stochastic nature of the proposed algorithm, different weighted independent sets might potentially be generated from the same input by running steps 3-11 can be run multiple times (see the condition in step 2).

 \begin{algorithm}[!t]
\caption{Random Independent Set Generator}\label{algo:greedypricing}
\begin{algorithmic}[1]
 \Require A Graph $\Graph = (\vertices, \edges), W$, a threshold weight $w_{min}$, and the maximum number of tries $t$.
 \Ensure A set of weighted independent sets.
\State $S\gets \emptyset$ 
\While{The maximum number of tries $t$ is not reached}
\State $\Graph' \gets G$
\State $IS\gets \emptyset$ 
\While{$\Graph'$ has vertices}
\State Randomly select a vertex in $\Graph'$
\State Add the selected vertex to $IS$
\State Remove the selected vertex and its neighbors from $\Graph'$.
\EndWhile
\If{IS was not generated before and its total weight is greater than $w_{min}$}
\State $S\gets S \cup IS$
\EndIf
\EndWhile
\State \textbf{return} $S$
\end{algorithmic}\end{algorithm}

 %\pagebreak

\bibliographystyle{IEEEtran}
\bibliography{bib}

\end{document}